\documentstyle[preprint,aps,eqsecnum]{revtex}
\begin{document}
\tighten

\def\eps{\varepsilon}

\draft
\title{Two Interacting Particles in a Random Potential: 
Mapping onto One Parameter Localization Theories without Interaction}

\author{Klaus Frahm$^{(1,2)}$, Axel M\"uller--Groeling$^{(1,3)}$ and 
        Jean--Louis Pichard$^{(1)}$}

\address{$^1$Service de Physique de l'\'Etat condens\'e,
        CEA--Saclay, 91191 Gif--sur--Yvette Cedex, France \\
        $^2$Instituut-Lorentz, University of Leiden, 
        P.O. Box 9506, 2300 RA  Leiden, The Netherlands \\
        $^3$Max--Planck--Institut f\"ur
        Kernphysik, Postfach 10 39 80, D--69029 Heidelberg, Germany
}


\maketitle

\begin{abstract}
 
We consider two models for a pair of interacting particles in a random
potential: (i) two particles with a Hubbard interaction in arbitrary
dimensions  and (ii) a strongly bound pair in one dimension.
Establishing suitable correpondences we demonstrate that
both cases can be described in terms familiar from theories of
{\it noninteracting} particles. In particular, these two cases are
shown to be controlled by a single scaling variable, namely the pair
conductance $g_2$. For an attractive or repulsive Hubbard interaction
and starting from a certain effective Hamiltonian we derive a
supersymmetric nonlinear $\sigma$ model. Its action turns out to be
closely related to the one found by Efetov for noninteracting
electrons in disordered metals. This enables us to describe the
diffusive motion of the particle pair on scales exceeding the
one--particle localization length $L_1$ and to discuss the
corresponding level statistics. For tightly bound pairs in one
dimension, on the other hand, we follow early work by Dorokhov and
exploit the analogy with the transfer matrix approach to  quasi--1d
conductors. Extending our study to $M$ particles we obtain a
$M$--particle localization length scaling like the $M$th power of the
one--particle localization length.

\end{abstract}

\narrowtext

\section{Introduction}

The concept of one parameter scaling 
pioneered by Thouless \cite{thouless} and generalized by 
Abrahams et al \cite{scaling1,scaling2} 
has been very fruitful in the understanding of the localization problem for 
noninteracting electrons in a random potential. 
The relevant scaling parameter 
is the size--dependent conductance $g(L)$ (in units of $2e^2/h$) determined 
by the so--called $\beta$--function $\beta(g)$ \cite{scaling1}. This 
$\beta$--function which contains all the relevant information 
has in principle to be calculated from a microscopic 
theory. However, the general field theoretical formulation in terms of a 
non linear $\sigma$ model \cite{wegner1,efetov} shows the universal 
behavior of $\beta(g)$ and that it only depends on the symmetry class and 
the space dimension. In $2+\eps$ dimensions the $\beta$--function 
is accessible by diagrammatic perturbation theory 
\cite{efetov,wegner2,hikami}, and in quasi--one dimension even ``exact'' 
solutions are known, implying exponential localization
\cite{larkin,efetov}. Furthermore, exact expressions for the first two 
moments of the conductance have been derived \cite{zirn_cond,mmz}. 

In one or quasi--one dimension there is, apart from the
$\sigma$ model, another 
powerful theory \cite{fokpla} to describe the scaling behavior of 
the conductance. In this approach the distribution of transmission 
eigenvalues is determined by a Fokker--Planck equation, the 
Dorokhov-Mello-Pereyra-Kumar (DMPK) equation, describing how the 
transmission properties evolve with the length of the wire. 
In the unitary symmetry class the entire distribution of transmission 
eigenvalues \cite{GUEcase} and the first two conductance moments 
\cite{frahm_let} are known for arbitrary length scales. 
Moreover, it has become clear recently that the DMPK equation and the
$\sigma$ model formulation of quasi--1d transport are mathematically
equivalent \cite{frahm_let,eq_rejaei,equiv}. In both approaches,
one--parameter scaling emerges in a natural way.

Not long ago, Shepelyansky \cite{shep1} studied the localization 
problem of {\em two interacting\/} particles (TIP). He considered 
a one--dimensional disordered potential and a local 
(Hubbard--like) repulsive or attractive interaction $U$. 
Assuming some generic properties of the interaction induced 
coupling matrix elements between localized one--electron product states, 
Shepelyansky found that certain TIP--states are extended on 
a scale given by the {\em two--particle} localization length 
$L_2\propto L_1^2$, which can be 
much larger than the one--particle localization 
length $L_1$. Using similar assumptions, Imry \cite{imry} was able to 
confirm Shepelyansky's original result for $L_2$ by using the Thouless 
\cite{thouless} block scaling picture. He introduced the notion 
of a {\em two--particle} conductance $g_2(L)$ and {\em assumed} 
that it obeys the same scaling behavior as the one--particle conductance. 
In subsequent work, the enhancement of the TIP localization length 
was confirmed by numerical studies for the 
transfer matrix \cite{fmgpw}, the TIP--Hamiltonian \cite{wmgpf} and 
the TIP--Green's function \cite{oppen1}. 

In this paper we show that both the $\sigma$ model and the Fokker--Planck 
approach can be succesfully applied to the TIP--localization problem, 
thus relating the TIP--problem to an equivalent problem without interactions. 
This allows us to identify a certain pair conductance $g_2$ 
associated with TIP--diffusion and TIP--localization and to {\em show} 
that $g_2$ is governed by the same scaling function as the 
one--electron conductance. Furthermore, we relate $g_2$ 
to the statistical properties of the TIP energy levels and to 
the transmission eigenvalues of the corresponding scattering 
problem. 

In section \ref{section:2} we consider, generalizing Shepelyansky, 
the TIP--problem in $d$ dimensions with a Hubbard--like repulsive or 
attractive interaction $U$. We first motivate an effective random 
matrix model that incorporates precisely the $2d$--dimensional 
lattice structure of the two--electron pair states. The only 
assumptions concern the statistical independence and the typical size 
of the interaction--induced coupling matrix elements between these states. 
The next step is to map the random matrix model onto 
a supersymmetric non linear $\sigma$ model. The corresponding 
action is after appropriate identifications formally 
identical with the action of Efetov's ``standard'' 
$\sigma$ model \cite{efetov} for disordered conductors. 
This allows us to determine in 
arbitrary dimensions the Breit--Wigner form for the local density of
states, the different frequency regimes for the diffusion of 
interaction--assisted pairs at scale $L>L_1$, and the two point 
correlation for the corresponding TIP--levels. We relate $g_2$ to 
a pair Thouless energy $E_c^{(2)}$ defined as the energy 
range for finding Wigner--Dyson level statistics. However, we also 
obtain a subtle modification with respect to the one--particle 
problem: the diffusion is logarithmically suppressed and 
the size of the interaction--assisted pairs increases 
logarithmically with time. These results are in agreement with recent 
findings of Borgonovi and Shepelyansky \cite{shep3,shep4} 
based on numerical and qualitative arguments. A summary of our main 
results has already been published in a separate publication 
\cite{fmgp} and we present here a comprehensive derivation of the 
different steps. 

In section \ref{section:3}, we consider a different TIP problem,
which was first studied by Dorokhov \cite{dorokhov}: 
a tightly bound pair of particles in a strictly $1d$ random potential. 
In analogy to quasi one--dimensional conductors, we treat this problem 
in terms of the DMPK equation. The scattering channels for the transfer matrix 
formulation are now provided by the internal quantum modes of the 
bound pair. The conductance $g_2$ is 
then related to the transmission eigenvalues corresponding to these 
scattering channels. To identify $g_2$ in this approach, we have to 
determine one microscopic length scale, namely the effective mean free path 
or equivalently the localization length as a function of the microscopic 
parameters. Again we find that the ``conductance'' $g_2$ 
for tightly bound pairs has the same scaling with system size 
as the conductance of non interacting electrons in quasi 
one--dimensional wires. 
Extending this approach to $M$ bound particles in one 
dimension, we obtain a delocalization effect which increases with 
$M$ as $L_1^M$ if the particles are 
forced to stay within a distance of order $L_1$ from each other.
The delocalization of certain $M$ particle states by 
a local interaction will be discussed in a greater generality 
elsewhere \cite{ip}.

\section{Two particles with local attractive or repulsive interactions 
in d-dimensions}

\label{section:2}

We consider two interacting electrons on a lattice in $d$ 
dimensions with site diagonal disorder and nearest neighbor hopping 
elements. We assume that all one electron eigenstates are localized with a 
typical localization length $L_1\gg 1$ (all length scales will be measured 
in units of the lattice constant). 
This requires  the disorder to be sufficiently weak in one or two
dimensions, while the disorder strength should be slightly above the
critical disorder in three dimensions.
The localized one electron states $\varphi_\rho(r)$ are labeled by an index 
$\rho$ describing the ``center'' of the state. We consider the case 
of symmetric two--electron wave functions $\psi(r_1,r_2)=\psi(r_2,r_1)$ 
(e.g. both electrons have opposite spin) and express the 
Hamilton operator in the basis of symmetrized products of one--electron  
states $|\rho_1 \rho_2>$, given by the wave function 
$[\varphi_{\rho_1}(r_1)\varphi_{\rho_2}(r_2)+
\varphi_{\rho_1}(r_2)\varphi_{\rho_2}(r_1)]/\sqrt{2}$:
\begin{equation}
\label{eq:1}
H=\sum_{\rho_1, \rho_2} (\eps_{\rho_1}+\eps_{\rho_2})
|\rho_1 \rho_2><\rho_1 \rho_2|+\sum_{\rho_1, \rho_2,\rho_3, \rho_4}
Q_{\rho_1 \rho_2 \rho_3 \rho_4}\,|\rho_1 \rho_2><\rho_3 \rho_4|\ ,
\end{equation}
where  $Q_{\rho_1 \rho_2 \rho_3 \rho_4}$ are the interaction matrix 
elements. The quantities
$\eps_\rho$ are the one--electron energies of the localized states 
$\varphi_\rho(r)$. For a local Hubbard interaction with interaction 
strength $U$ 
\begin{equation}
\label{eq:2}
H_{int}=U\sum_r |rr><rr|
\end{equation}
we have for instance
\begin{equation}
\label{eq:3}
Q_{\rho_1 \rho_2 \rho_3 \rho_4}=<\rho_1 \rho_2|\,H_{int}\,|\rho_3 \rho_4>
=2U\sum_r \varphi_{\rho_1}(r)\,\varphi_{\rho_2}(r)\,\varphi_{\rho_3}(r)\,
\varphi_{\rho_4}(r)\quad.
\end{equation}
These interaction matrix elements become exponentially small if two of 
the ``positions'' $\rho_j$ are far away on a scale much larger than the 
one--electron localization length. Shepelyansky divided \cite{shep1} 
(in one dimension) the problem described by the Hamiltonian (\ref{eq:1}) 
into two blocks: a diagonal one corresponding to matrix elements 
$<\rho_1 \rho_2 | H | \rho_3 \rho_4>$, where 
at least two of the positions $\rho_j$ are separated by more than $L_1$,
and a random band matrix with a 
superimposed strong diagonal (SBRM), where every $\rho_j$ is at less 
than $L_1$ from the three others. The ensuing mathematical random matrix
model has been investigated in \cite{shep2,fyodorov2,framg1}.
The two crucial simplifications
for this picture 
are first to neglect all exponentially small interaction matrix elements 
$Q_{\rho_1 \rho_2 \rho_3 \rho_4}$ ( off-diagonal terms of the first block,  
off-diagonal terms outside the band of the second block, coupling terms 
between the two blocks) and second to assume that the remaining coupling 
elements (\ref{eq:3}) are independent random variables with typically 
the same size. Using the ansatz
\begin{equation}
\label{eq:4}
\varphi_\rho(r)\sim \frac{1}{(L_1)^{d/2}}\,e^{-|r-\rho|/ L_1}
\,a_\rho(r)\quad,
\end{equation}
and applying a central limit theorem argument in (\ref{eq:3}), one may 
estimate \cite{shep1,imry} the coupling strength between well coupled 
product states of the SBRM-block to be
\begin{equation}
\label{eq:5}
\left\langle Q_{\rho_1 \rho_2 \rho_3 \rho_4}^2\right\rangle \sim 
U^2 L_1^{-3d}\quad.
\end{equation}
In (\ref{eq:4}), we assume for simplicity that $a_\rho(r)$ is 
a random variable with $\langle a_\rho(r)\rangle =0$ and $\langle a_\rho(r)
a_{\tilde\rho}(\tilde r)\rangle =\delta_{\rho \tilde\rho}\,
\delta_{r \tilde r}$. 
 
\subsection{ Effective Random Matrix Hamiltonian}

In this study, we will consider an effective random matrix Hamiltonian 
(ERMH), which relies only on the second of 
the above--mentioned simplifications. In other 
words, {\em all} product states $|\rho_1 \rho_2>$ will be taken into 
account, and we keep the total Hamiltonian, and not only the SBRM sub-block 
where the states are strongly affected by the interaction.  However,
the coupling matrix elements (\ref{eq:3}) of the total 
Hamiltonian matrix will still be assumed to be 
independent (Gaussian) variables. This latter point is indeed quite a 
serious simplification since the correlations between different 
coupling matrix elements will be neglected. In particular, the exact 
operator identity $H_{int}^2=U H_{int}$ [see (\ref{eq:2})] is 
violated. However, we believe that our 
ERMH will nevertheless be quite
realistic due to the strongly fluctuating diagonal 
elements, which tend to eliminate the effect of the correlations. 

We consider a $2d$--dimensional configuration space (lattice) with 
two coordinate vectors $R\equiv \rho_1+\rho_2$ and $j\equiv(\rho_1-\rho_2)/2$ 
corresponding to twice the center of mass and half the distance of the two 
electrons. This choice ensures that the components of these 
coordinates differ for adjacent lattice sites just by $+1$ or $-1$ if 
the corresponding other coordinate is kept constant. The ERMH consists of 
two parts
\begin{equation}
\label{eq:6}
{\cal H}=\hat\eta+\hat \zeta \ ,
\end{equation}
where the
\begin{equation}
\label{eq:7}
(\hat\eta)_{R \tilde R}^{\,j\, \tilde j}=\eta_R^j\ \delta_{R \tilde R}
\,\delta_{j \tilde j}
\end{equation}
correspond to the diagonal entries $\eps_{\rho_1}+\eps_{\rho_2}$ of 
(\ref{eq:1}). For simplicity, we choose for the $\eta_R^j$ independent 
random variables with the distribution function $\rho_0(\eta)$. We have 
typically $\rho_0(\eta)\simeq 1/(2W_b)$ for $|\eta|\le W_b$ where $W_b$ 
is twice the bandwidth of the disordered one--electron Hamiltonian. The 
operator $\hat\zeta$ corresponds to the interaction matrix elements 
(\ref{eq:3}). The matrix elements are independent Gaussian random 
variables with zero mean and variance
\begin{equation}
\label{eq:8}
\left \langle \left(\hat\zeta_{R \tilde R}^{\,j\, \tilde j}\right)^2\right
\rangle =\frac{1}{2}(1+\delta_{R \tilde R}\,\delta_{j \tilde j})
\,a(|R-\tilde R|)\, v(j)\,v(\tilde j)\quad.
\end{equation}
The smooth functions
$a(|R-\tilde R|)$ and $v(j)$ decay exponentially 
on the scale $L_1$. We assume the typical behavior: 
\begin{equation}
\label{eq:9}
a(|R-\tilde R|)\sim 
\left\{\begin{array}{ll}
U^2 L_1^{-3d} & ,\ |R-\tilde R|\lesssim L_1\ , 
\phantom{\Big|}\\
U^2 L_1^{-3d} e^{-2|R-\tilde R|/L_1} & 
,\ |R-\tilde R|\gg L_1\ , \phantom{\Big|}\\
\end{array}\right.
\end{equation}
and
\begin{equation}
\label{eq:10}
v(j)\sim\left\{
\begin{array}{ll}
1 & ,\ |j|\lesssim L_1\ , \phantom{\Big|}\\
e^{-4|j|/L_1} & ,\ |j|\gg L_1\ , \phantom{\Big|}\\
\end{array}\right.
\end{equation}
which is justified by Eqs. (\ref{eq:3}) and (\ref{eq:4}). The function 
$a(|R-\tilde R|)$ describes how the coupling strength decreases 
exponentially as the distance of the centers of mass increases, while
$v(j)$ describes how the ``size'' of the product states reduces the 
coupling. 

In Eqs. (\ref{eq:9}), (\ref{eq:10}), we have applied the simplified 
estimate (\ref{eq:5}) for the typical strength of the interaction 
coupling matrix elements. However, the presented random matrix 
model and its subsequent treatment is given in terms of the general 
functions $a(R)$ and $v(j)$. It is straightforward to use different, 
more ``realistic'', estimates instead of (\ref{eq:5}), e.g. to take 
effects of ballistic or diffusive motion for length scales $L\lesssim L_1$ 
into account.

\subsection{Nonlinear $\sigma$ Model}

Our goal is to investigate the spectral statistics, the transport and the 
localization properties of the ERMH (\ref{eq:6}) applying the supersymmetry
technique \cite{efetov,vwz}. In the following, we choose a particular 
realization of the diagonal elements $\eta_R^j$ and perform the ensemble 
average only with respect to $\hat\zeta$. This particular average is 
denoted by $\langle \cdots\rangle_\zeta$. 
We consider the generating functional
\begin{eqnarray}
\nonumber
F(J) & = & 
\left\langle \int D\psi\ \exp\left[\frac{i}{2}\bar\psi
\Bigl( E-{\cal H} +(\frac{\omega}{2}+i\eps)\Lambda+J\Bigr)\psi\right]
\right\rangle_\zeta\\
\label{eq:11}
& = & 
\left\langle \mbox{sdet}^{-1/2}
\Bigl( E-{\cal H} +(\frac{\omega}{2}+i\eps)\Lambda+J\Bigr)
\right\rangle_\zeta\quad.
\end{eqnarray}
Here, $\psi$ is a supervector with components $\psi_j(R)$, which are 
8--dimensional supervectors with entries $z_1,\bar z_1,\chi_1,\bar\chi_1,
z_2,\bar z_2,\chi_2,\bar\chi_2$, where $z_\nu$ ($\chi_\nu$) are complex 
bosonic (fermionic) variables. 
The diagonal matrix $\Lambda = {\rm diag}({\bf 1}_4,-{\bf 1}_4)$
describes the grading
into advanced and retarded Green's functions, $\omega$ is a 
frequency, $J$ is a source matrix, and $\bar\psi$ is given by $\bar\psi=
\psi^\dagger \Lambda$. For the graded determinant (``sdet'') and for the
graded trace (``str'') we use the convention that
$\mbox{str}(A)=\mbox{tr}(A_{BB})-\mbox{tr}(A_{FF})$ and 
$\mbox{sdet}(A)=\exp(\mbox{str}\,\ln(A))$, where $A_{BB}$ ($A_{FF}$) is the 
boson--boson (fermion--fermion) block of the supermatrix $A$. 

Using suitable choices \cite{vwz,iwz} for $J$ and taking derivatives of $F(J)$ 
with respect to $J$ at $J=0$, one can obtain moments of the Green's function. 
These contain all the informations about the spectral statistics and 
transport properties we are interested in. 
Our strategy is the following: We derive a nonlinear $\sigma$ model
from the functional $F(J)$ and discuss its physical implications by
comparing our result with the $\sigma$ model description of
noninteracting electrons derived by Efetov \cite{efetov}.

The ensemble average with respect to $\hat\zeta$ is readily performed and 
results in
\begin{equation}
\label{eq:12}
F(J)=\int D\psi\ \exp\left[\frac{i}{2}\bar\psi
\Bigl( E-\hat\eta +(\frac{\omega}{2}+i\eps)\Lambda+J\Bigr)\psi
-\frac{1}{8}\sum_{R,\tilde R} a(|R-\tilde R|)\,\mbox{str} 
[K(R)\, K(\tilde R)]
\right]\ ,
\end{equation}
where $K(R)$ are $8\times 8$ supermatrices given by
\begin{equation}
\label{eq:13}
K(R)=\sum_j v(j)\,\psi_j(R)\,\bar\psi_j(R)\quad.
\end{equation}
We see that the statistical properties (\ref{eq:8}) of $\hat\zeta$ 
allow for a convenient decoupling into the center of mass 
coordinates $R$ and the relative coordinates $j$. The quartic term in 
(\ref{eq:12}) is decoupled in a standard way \cite{efetov,vwz} by a 
Hubbard-Stratonovich transformation. This introduces an integral over a 
field of $8\times 8$ supermatrices $\sigma(R)$ with the same symmetries as 
$K(R)$, and we get
\begin{eqnarray}
\nonumber
F(J)&=&\int D\sigma \exp(-{\cal L}_1[\sigma])\quad,\\
\label{eq:14}
\tilde{\cal L}_2[\sigma]&=&\frac{1}{8}\sum_{R,\tilde R} (A^{-1})_{R \tilde R}
\ \mbox{str} [\sigma(R)\, \sigma(\tilde R)]
+\frac{1}{2} \mbox{str}_{(jR)}\ \ln \Bigl( {\textstyle 
E-\hat\eta +(\frac{\omega}{2}
+i\eps)\Lambda+J+\frac{1}{2}\hat\sigma}\Bigr)\quad.
\end{eqnarray}
Here, $A^{-1}$ is the inverse of the matrix 
$A_{R\tilde R}=a(|R-\tilde R|)$ and $\hat\sigma$ is given by
\begin{equation}
\label{eq:15}
(\hat\sigma)_{R\tilde R}^{\,j\,\tilde j}=v(j)\,\sigma(R)
\,\delta_{R \tilde R}\,\delta_{j \tilde j} \ .
\end{equation}
The supertrace (``$\mbox{str}_{(jR)}$'') appearing in (\ref{eq:14}) 
extends over the full $8L^{2d}$--dimensional supervector space, where
$L$ is the system size assumed to be much larger than $L_1$. 
To proceed, we apply as usual a saddle point approximation,
which is valid in the limit $L_1\gg 1$. The 
corresponding saddle point equations $\delta {\cal L}_1/\delta \sigma(R)=0$ 
can be written in the form (for $J=0$ and $\omega=0$)
\begin{equation}
\label{eq:16}
\sigma(R)=-\sum_{\tilde R} a(|R-\tilde R|) \sum_j v(j)
\frac{1}{E-\eta_{\tilde R}^j+i\eps\Lambda+\frac{1}{2}v(j)\sigma(\tilde R)}
\quad.
\end{equation}
The sums extend over $L_1^{2d}$ (contributing) terms and, applying a central 
limit theorem argument, we may replace the particular values of 
$\eta_{\tilde R}^j$ by an average with respect to the distribution 
$\rho_0(\eta)$ (see below for a more precise justification). Then 
(\ref{eq:16}) can be solved by a homogeneous 
(i.e. $R$--independent) ansatz $\sigma(R) =
\Gamma_0+i\Gamma_1\Lambda$, where $\Gamma_0,\Gamma_1$ are determined by 
the implicit equation
\begin{eqnarray}
\label{eq:17}
\Gamma_0+i\Gamma_1\Lambda & = & -B_0\sum_j v(j) \int d\eta\ \rho_0(\eta)
\,\frac{1}{E-\eta+i\eps\Lambda+\frac{1}{2}v(j)(\Gamma_0+i\Gamma_1\Lambda)}
\quad,\\
\label{eq:18}
B_0 & = & \sum_R a(|R|)\quad.
\end{eqnarray}
In the limit $L_1\gg 1$, $\Gamma_0$ and $\Gamma_1$ are much smaller than the 
typical integration range $2W_b$ for $\eta$ and we find
\begin{eqnarray}
\label{eq:19}
\Gamma_0 & \simeq & -B_0 S_1\,{\cal P}\int d\eta\ \rho_0(\eta) 
\frac{1}{E-\eta}\quad,\\
\label{eq:20}
\Gamma_1 & \simeq & \pi B_0 S_1\,\rho_0(E) \ ,
\end{eqnarray}
with $S_\nu=\sum_j [v(j)]^\nu\sim L_1^d$ and ${\cal P}\int d\eta\ 
(\cdots)$ a principal value integral. 
From Eqs. (\ref{eq:9}) and (\ref{eq:10}) we find the estimate
\begin{equation}
\label{eq:21}
\Gamma_1\sim\frac{U^2}{W_b}\,L_1^{-d}\ll W_b
\end{equation}
if $U$ and $W_b$ are of the same order of magnitude. The imaginary part of 
the r.h.s. of (\ref{eq:16}) is a sum of Lorentzians with a typical width 
$\Gamma_1 v(j)$. The most important contributions in this sum 
arise from $|j|\lesssim L_1$ and $|R-\tilde R|\lesssim L_1$ such that 
$v(j)\sim 1$. The maxima $\eta_R^j$ of the Lorentzians have thus a typical 
distance $\sim 2 W_b L_1^{-2d}\ll \Gamma_1$ and we are in the regime of well 
overlapping resonances. Therefore it is indeed allowed to replace 
(\ref{eq:16}) by its average with respect to $\eta_{\tilde R}^j$. 
 As usual, the saddle point $\Gamma_0+i\Gamma_1\Lambda$ is not unique and 
we have to consider a complete saddle point manifold $\Gamma_0+i\Gamma_1 Q$ 
with $Q=T^{-1}\Lambda T$, $T$ being an element of a certain supermatrix 
group 
\cite{efetov,vwz}. The supermatrix $Q$ fulfills the nonlinear constraint 
$Q^2=1$ and is in our case a member of the orthogonal $\sigma$ model space. 
In the following, we replace $\sigma(R)$ by $\Gamma_0+i\Gamma_1 Q(R)$ where 
each $Q(R)$ fulfills this constraint but may slowly vary in $R$-space. The 
matrix $A_{R\tilde R}=a(|R-\tilde R|)$ has eigenvalues 
$\tilde a(k)=\sum_R e^{ikR}\,a(|R|)$ and the eigenvalues of $A^{-1}$ 
are just given by
\begin{equation}
\label{eq:22}
\frac{1}{\tilde a(k)}\simeq B_0+k^2 \frac{B_2}{B_0^2}\quad,\quad
B_2=\frac{1}{2d}\sum_R R^2 a(|R|)\quad.
\end{equation}
Here we have applied an expansion with respect to small momenta $k$ since 
mainly the {\em slow} modes will give the relevant contributions
\cite{efetov,fyodorov}. In analogy to
\cite{fyodorov}, we evaluate the first term of the action 
(\ref{eq:14}) 
by performing a Fourier transformation, applying (\ref{eq:22}) and
afterwards transforming back to position space.
Since $Q(R)$ varies slowly with $R$, 
we may go over to a continuum limit and obtain the following expression for 
the functional $F(J)$:
\begin{eqnarray}
F(J) & = & \int DQ\, \exp(-{\cal L}_2[Q])\quad,
\nonumber\\
{\cal L}_2[Q] & = & -\frac{\Gamma_1^2 B_2}{8 B_0^2}\int dR\ \mbox{str}
\left([\nabla_R\, Q(R)]^2\right)
+\frac{1}{2} \mbox{str}_{(jR)}\ \ln \Bigl( {\textstyle 
E-\hat\eta +(\frac{\omega}{2}
+i\eps)\Lambda+J+\frac{1}{2}\hat\sigma}\Bigr)\quad,
\nonumber\\
\label{eq:23}
(\hat\sigma)_{R\tilde R}^{\,j\,\tilde j} & = & v(j)\,
[\Gamma_0+i\Gamma_1 Q(R)]
\,\delta_{R \tilde R}\,\delta_{j \tilde j}\quad.
\end{eqnarray}
For convenience we have kept the discrete notation for $R$ in the
second term of the action  ${\cal L}_2$. Expression (\ref{eq:23})
defines the nonlinear $\sigma$ model suitable to describe the ERMH.

\subsection{Breit-Wigner Form for the Local Density of States}

As a first application, we determine the local density of states, 
\begin{equation}
\label{eq:24}
\rho_R^j(E)=-\frac{1}{\pi}\lim_{\eps\to 0+}
\mbox{Im}\left\langle<Rj|(E-{\cal H}+i\eps)^{-1}|Rj>
\right\rangle_\zeta\quad.
\end{equation}
This quantity 
can be calculated using small $4\times 4$ supermatrices $\sigma(R)$ in 
(\ref{eq:14}) and replacing $\Lambda$ by 1. Then only the saddle point 
$\Gamma_0+i\Gamma_1$ (and not a full saddle point manifold)
contributes. Choosing for $J$ a supermatrix that has 
its only nonvanishing diagonal entry $x L_g$  at site $(j,R)$ 
($L_g$ is a diagonal supermatrix with entries $+1$ and $-1$ in the
boson--boson and fermion--fermion blocks, respectively), 
one finds
\begin{equation}
\label{eq:25}
\rho_R^j(E)=\frac{1}{\pi}\ \frac{\frac{1}{2}\Gamma_1\, v(j)}
{[E-\eta_R^j+\frac{1}{2}\Gamma_0\, v(j)]^2 + [\frac{1}{2} \Gamma_1\,
v(j)]^2}\quad.
\end{equation}
This is a Breit--Wigner function with the width $\frac{1}{2}\Gamma_1\,v(j)$. 
The same result can also be obtained by a standard diagrammatic expansion in 
powers of $\hat\zeta$. The complex self--energy is then given by
by $\Sigma_R^j=-\frac{1}{2}(\Gamma_0+i\Gamma_1)\,v(j)$ 
with $\Gamma_0$ and $\Gamma_1$ determined by 
(\ref{eq:17}) and (\ref{eq:18}).

To interpret (\ref{eq:25}) we note that the noninteracting product
states with eigenenergies $\eta_R^j$ acquire a finite width $\Gamma_1
v(j)$ due to the interaction. This width is of the order $\Gamma_1$
for $|j| \le L_1$ and decreases esponentially as 
$\Gamma_1\,e^{-4|j|/L_1}$ for $|j|\gg L_1$. This means that those
states $\eta_R^j$ constructed from well separated one--electron states
($|j| \gg L_1$) constitute
good ``quasi-levels'' with exponentially large life times.  
We may consider a critical pair size $L_c$ such that the 
pair states with $2|j|\lesssim L_c$ in a system of size $L_{\rm sys}$ 
are still reasonably well coupled. For this, 
the effective level spacing $\Delta_{\rm eff}=W_b\,(L_c\,L_{\rm sys})^{-d}$ 
has to be compared with the smallest possible level width 
$\Gamma_1\,e^{-2L_c/L_1}$. This yields the rough estimate
$L_c\sim L_1\ln L_{\rm sys}$. 
The $\eta_R^j$ with $|j|>L_c$ contribute to a discrete 
point spectrum whereas the states with $|j|\lesssim L_c$ are well mixed 
and form effective pairs of maximal size 
$L_c\sim L_1\ln L_{\rm sys}$.

\subsection{TIP-conductance $g_2$ at scale $L > L_1$ }

We now want to describe the dynamics of these pairs in terms of the 
effective $\sigma$ model (\ref{eq:23}) with the action ${\cal L}_2[Q]$. 
In order to understand the relation with a {\em disordered metal}, we 
recall a few results for the corresponding $\sigma$ model description 
given by Efetov \cite{efetov} and characterized by the action
\begin{equation}
\label{eq:26}
{\cal L}_{met}[Q]=-\frac{\pi}{8}\,\nu{\cal D}\int dr\ \mbox{str}
\left((\nabla Q)^2\right)+\frac{\pi\nu}{2}(-i\frac{\omega}{2}+\eps)
\int dr\ \mbox{str}(Q(r)\Lambda)\quad.
\end{equation}
Here ${\cal D}$ denotes the classical diffusion constant and $\nu$ the local 
density of states. Efetov derived \cite{efetov} this action for frequencies 
$|\omega|\ll 1/\tau$ where $\tau$ is the elastic scattering time due to the 
disorder. The action ${\cal L}_{met}[Q]$ therefore describes the 
{\em quantum} diffusion for time scales larger than $\tau$. 
The nontrivial quantum properties are due to the nonlinear 
constraint $Q^2=1$ of the supermatrix $Q(r)$. One may parameterize 
\cite{efetov} $Q$ by $Q=\Lambda\frac{1+iP}{1-iP}$ where $P$ is a supermatrix 
with only offdiagonal entries in the $\Lambda$--grading,
$\Lambda P\Lambda=-P$.
Expanding the action (\ref{eq:26}) with respect to $P$, Efetov showed that 
a perturbative evaluation of the $\sigma$ model is completely equivalent 
to the diagrammatic theory of disordered metals in terms of (interacting) 
Diffuson and Cooperon modes. The leading order corresponds to the 
{\em classical} diffusion and the higher order terms yield the first 
quantum corrections. Furthermore, Efetov found that the action 
(\ref{eq:26}) leads to the well known level statistics \cite{mehta} 
of random matrix theory if the considered range of energies is smaller 
than the Thouless energy $E_c={\cal D}/L^2$ (i.e. $|\omega|\ll E_c$).
For frequencies $\omega$ such that 
$E_c\ll|\omega|\ll 1/\tau$ and in the metallic regime $E_c\gg \Delta$
($\Delta$ denotes the level spacing), the corrections to 
the universal Wigner-Dyson statistics have been calculated by Altshuler 
and Shklovskii \cite{alt1}. 
In one dimension, the $\sigma$ model (\ref{eq:26}) describes a (thick) wire.
All the wave functions are localized \cite{larkin,efetov} with a 
localization length $\xi=2\pi \nu{\cal D}$ (for $L\gg \xi$). 

Comparing the first terms of the actions of (\ref{eq:23}) and 
(\ref{eq:26}), one gets for two interacting particles 
\begin{equation}
\label{eq:27}
\nu_{\rm eff}{\cal D}_{\rm eff}=\frac{1}{\pi}\,\frac{\Gamma_1^2\,B_2}{B_0^2}=
\pi\rho_0(E)^2 S_1^2 B_2\sim \frac{U^2}{W_b^2} L_1^2, 
\end{equation}
where
the second identity results from Eq. (\ref{eq:20}). We have 
introduced an effective (pair) diffusion constant ${\cal D}_{\rm eff}$ and an 
effective local density of 
states $\nu_{\rm eff}$. For (quasi) one--dimensional 
systems the quantity in (\ref{eq:27}) is proportional to the localization 
length $L_2$ for the center of mass coordinate for the pairs and we recover 
Shepelyansky's result \cite{shep1}: $L_2\sim (U^2/W_b^2)\,L_1^2$.
We reiterate that we have taken {\it all}, also the weakly coupled,
product states into account. It has been suggested \cite{shep2} that
these weakly coupled states lead to a logarithmic correction,
$L_2\sim L_1^2/\ln L_1$. From our analytical considerations we find no
evidence for such a behavior.

In order to extract the diffusion constant 
${\cal D}_{\rm eff}$ from (\ref{eq:27}),
we still need a reasonable estimate for the effective 
local density of (product) states $\nu_{\rm eff}$ 
that contribute to the diffusion.
Naively, one could consider the effective size $L_c\sim L_1\ln 
L_{\rm sys}$ of 
the pair states, which we estimated in the above discussion subsequent to 
Eq. (\ref{eq:25}). This would yield the estimates
\begin{equation}
\label{eq:28}
\nu_{\rm eff}\sim \rho_0(E)\, L_c^d\sim \rho_0(E)\,L_1^d(\ln L_{\rm sys})^d
\quad\Rightarrow\quad {\cal D}_{\rm eff}\sim \frac{U^2}{W_b} L_1^{2-d}\,
(\ln L_{\rm sys})^{-d}\quad.
\end{equation}
However, this simple picture assumes that all product states up to a size of 
$L_c$ contribute equally to the diffusion. 
Obviously, this is not the case.
Similarly to the frequency 
condition $|\omega| <1/\tau$ for a disordered metal, we have to require 
$|\omega|<1/\tau_c$ to justify (\ref{eq:28}),
where $1/\tau_c=\Gamma_1 v(L_c)=\Delta_{\rm eff}$ 
is the effective level spacing of the well coupled product states,
which is by definition
equal to the inverse life time of pairs of size $L_c$. We recall that 
$1/\tau_c=W_b L_c^{-2d}\sim W_b(L_1\ln L_{\rm sys})^{-2d}$ 
is much smaller than 
the energy scale $\Gamma_1\sim (U^2/W_b) L_1^{-d}$, which is the 
inverse life time of the {\em best} coupled states (of size 
$|j|\lesssim L_1=L_c/\ln L_{\rm sys}$). For frequencies between $1/\tau_c$ and 
$\Gamma_1$, we thus expect that the number $N(\omega)$ of contributing 
states depends on $\omega$ via:
\begin{equation}
\label{eq:29}
N(\omega)=[L_{\rm eff}(\omega)]^d\quad,\quad
L_{\rm eff}(\omega)=\frac{L_1}{4}\ln\left(\frac{\Gamma_1}{|\omega|}\right)
\ ,
\end{equation}
where $L_{\rm eff}(\omega)$ is the pair size 
such that the corresponding inverse 
life time $\Gamma_1\, v\Bigl(L_{\rm eff}(\omega)\Bigr)$ equals $|\omega|$. Eq. 
(\ref{eq:29}) results in a frequency dependent density of states and 
diffusion constant,
\begin{equation}
\label{eq:30}
\nu_{\rm eff}(\omega)\sim \rho_0(E)\,\left(L_1\ln\Bigl(\Gamma_1/|\omega|\Bigr)
\right)^d\quad\Rightarrow\quad {\cal D}_{\rm eff}(\omega)\sim \frac{U^2}{W_b}
L_1^{2-d}\left(\ln\Bigl(\Gamma_1/|\omega|\Bigr)
\right)^{-d}\quad.
\end{equation}
This estimate is valid for $1/\tau_c\lesssim |\omega|\lesssim 
\Gamma_1$, whereas for $|\omega|\lesssim 1/\tau_c$ Eq. (\ref{eq:28}) applies. 
The estimates (\ref{eq:28}) and (\ref{eq:30}), 
based on the simple but physical 
life time argument, can also 
be obtained in a more formal way directly from 
the $\sigma$ model action (\ref{eq:23}). For this we rewrite ${\cal L}_2[Q]$ 
(at vanishing source term $J=0$) as
\begin{equation}
\label{eq:31}
{\cal L}_2[Q]=-\frac{\pi}{8}\nu_{\rm eff}\,{\cal D}_{\rm eff}
\int dR\ \mbox{str}\left((\nabla Q)^2\right)+
\int dR\ f_R(Q(R)) \ ,
\end{equation}
where 
\begin{equation}
\label{eq:32}
\label{EQ:32}
f_R(Q)=\frac{1}{2}\sum_j \mbox{str}\,\ln \Bigl( {\textstyle 
E-\eta_R^j +(\frac{\omega}{2}
+i\eps)\Lambda+\frac{1}{2}v(j)\,(\Gamma_0+i\Gamma_1 Q)}\Bigr)
\end{equation}
is some effective $Q$--potential for the $\sigma$ model. 
In Appendix \ref{app:klaus_a}, 
this expression for the effective potential is simplified to give
\begin{equation}
\label{eq:33}
f_R(Q)\simeq -i\frac{\pi}{4}\omega\ h\left(\frac{\Gamma_1}{\omega}\right)
\,\rho_0(E)\ \mbox{str}(Q\Lambda) \ ,
\end{equation}
where $h(\Gamma_1/\omega)$ is defined in Eq. (\ref{eq:a6}) of 
Appendix \ref{app:klaus_a} and has the limiting behavior described 
in (\ref{eq:a7}--\ref{eq:a9}). 
Comparing (\ref{eq:33}) with the second term in the action 
(\ref{eq:26}), we obtain a complex effective density of states,
\begin{equation}
\label{eq:34}
\nu_{\rm eff}(\omega)=\rho_0(E)\ h\left(\frac{\Gamma_1}{\omega}\right)\quad.
\end{equation}
For the two cases (\ref{eq:a8}) and (\ref{eq:a9}) we directly recover the 
estimates (\ref{eq:30}) and (\ref{eq:28}), respectively. On the other hand, 
the frequency range $|\omega| \gg \Gamma_1$ correponds to
\begin{equation}
\label{eq:35}
f_R(Q)\simeq \frac{\pi}{4}\,\Gamma_1\ \mbox{str}(Q\Lambda)\ ,
\end{equation}
which means that the action ${\cal L}_2[Q]$ no longer depends on $\omega$.

To summarize, we have found that for frequencies smaller than $\Gamma_1$, the 
$\sigma$ model action ${\cal L}_2[Q]$ can be written in a similar form as 
the action ${\cal L}_{met}[Q]$ for a disordered metal,
\begin{equation}
\label{eq:39}
{\cal L}_2[Q]=-\frac{\pi}{8}\,\nu_{\rm eff}(\omega)\,{\cal D}_{\rm eff}(\omega)
\int dr\ \mbox{str}
\left((\nabla Q)^2\right)+\frac{\pi\nu_{\rm eff}(\omega)}{2}
(-i\frac{\omega}{2}+\eps)
\int dr\ \mbox{str}(Q(r)\Lambda) \ .
\end{equation}
It was necessary to introduce both a frequency dependent diffusion
constant ${\cal D}_{\rm eff}(\omega)$
and a frequency dependent density of states $\nu_{\rm eff}(\omega)$,
\begin{eqnarray}
{\cal D}_{\rm eff}(\omega) & \sim & 
\left\{\begin{array}{ccl}
\frac{U^2}{W_b}\,L_1^{2-d}\,(\ln L_{\rm sys})^{-d} & \quad,\quad & 
|\omega|\lesssim {1/\tau_c} \\
&&\\
\frac{U^2}{W_b}\,L_1^{2-d}\,\left(\ln(\frac{\Gamma_1}{|\omega|})\right)^{-d} 
& \quad,\quad & 
{1/\tau_c}\lesssim |\omega|\ll \Gamma_1 \\
\end{array}\right.
\label{eq:40}\\
\nonumber\\
\nu_{\rm eff}(\omega) & = & \frac{\sigma_{\rm eff}}{{\cal D}_{\rm eff}(\omega)}
\quad,\quad \sigma_{\rm eff}=\frac{\Gamma_1^2 B_2}{\pi B_0^2}\sim 
\frac{U^2}{W_b^2}\,L_1^2\quad.
\label{eq:41}
\end{eqnarray}
Here, we have defined in analogy to a disordered metal via the Einstein 
relation a ``pair conductivity'' 
$\sigma_{\rm eff}$ which does {\em not} depend 
on frequency. This determines the TIP-conductance $g_2$ mentioned 
in the introduction through the usual relation between conductance 
and conductivity: $g_2 = \sigma_{\rm eff} L_{\rm sys}^{d-2}$. 
The analogy with a 
disordered metal allows us to 
draw some very interesting and far reaching conclusions. First, we can 
identify the coupling constant $\frac{\pi}{8}\nu_{\rm eff}\,{\cal D}_{\rm eff}=
\frac{\pi}{8}\sigma_{\rm eff}$ in ${\cal L}_2[Q]$ as a universal scaling
parameter. The corresponding scaling function is precisely the same as 
that of a disordered metal provided the latter is described by the 
``standard'' $\sigma$ model (\ref{eq:26}). In particular, the perturbative 
evaluation of the $\beta$--function in $2+\eps$ dimensions 
\cite{efetov,wegner2,hikami} is also valid for our problem of 
{\em diffusing} or {\em localized} electron pairs (the second term in 
${\cal L}_2[Q]$ 
containing the supertrace $\mbox{str}(Q\Lambda)$ is not affected under the 
renormalization \cite{efetov}). In one dimension, 
we are able to directly identify the pair localization length
$L_2\sim \sigma_{\rm eff}\sim (U^2/W_b^2) L_1^2$. These remarks
rigorously justify Imry's application \cite{imry} of the Thouless
scaling picture \cite{thouless}.

\subsection{ TIP-Dynamics for $ L_1 << L \leq L_2$}

A second conclusion concerns the pair--dynamics in the metallic (or 
diffusive) regime ($L\lesssim L_2$ for $d=1,2$) where the $\sigma$ 
model can be treated perturbatively (cp. Ref. \cite{efetov}). The relevant 
diffusion propagator ${\cal R}(q,\omega)=[\sigma_{\rm eff}\, q^2 
-i\omega\nu_{\rm eff}(\omega)]^{-1}=\{\nu_{\rm eff}(\omega)
[{\cal D}_{\rm eff}(\omega)\,
q^2 -i\omega]\}^{-1}$ now contains 
the frequency dependent diffusion constant (\ref{eq:40}) and density of 
states (\ref{eq:41}). This gives rise to some subtle modifications of standard 
diffusion. The precise calculation of the diffusion propagator 
$\tilde{\cal R}(R,t)=(2\pi)^{-(d+1)}\int d\omega\int dq\ {\cal R}(q,\omega)
\,e^{i(qR-\omega t)}$ in position and time space seems to be rather difficult 
due to the various frequency 
dependencies involved. However, in a 
qualitative approximation, we may replace $\omega$ by $1/t$. 
This shows that
the pairs propagate according to
\begin{equation}
\label{eq:42}
\langle R^2(t)\rangle \sim {\cal D}_{\rm eff}(1/t)\,t\sim 
\left\{\begin{array}{ll}
L_1^2 \Bigl(1+\alpha\Gamma_1\,t+\cdots)=L_1^2+\tilde\alpha D_0 t +\cdots
&,\ t\ll\Gamma_1^{-1}\phantom{\Big|}\\
D_0\,[\ln(\Gamma_1 t)]^{-d}\,t & ,\ \Gamma_1^{-1}\ll t
\lesssim \tau_c \phantom{\Big|}\\
D_0\,[\ln L_{\rm sys}]^{-d}\,t & ,\ \tau_c\lesssim t \phantom{\Big|}\\
\end{array}\right.
\end{equation}
We have used (see \ref{eq:40}) the notation 
$D_0=(U^2/W_b)\,L_1^{2-d}$, and $\alpha$ and $\tilde\alpha$ are numerical 
constants of the order of unity. 
For very small time scales $t\ll\Gamma_1^{-1}$ 
the dynamics is not diffusive. The corresponding estimate
given in (\ref{eq:42}) for these time scales may formally be 
obtained by a naive combination of (\ref{eq:a7}), (\ref{eq:34}) and 
(\ref{eq:41}), which is strictly speaking not justfied. However, a more 
sophisticated analysis for the classical propagator $\tilde{\cal R}(R,t)$ 
shows that this estimate is indeed correct. The physical interpretation 
of (\ref{eq:42}) is that the electron pairs are immediately 
(on time scales which are not resolved by the ERMH) delocalized 
on a scale $L_1$, due to interaction induced random hoppings. 
Then a diffusion with diffusion constant $D_0$ sets in. 
Eventually the diffusion constant is suppressed
by the time dependent factor 
$[\ln(\Gamma_1 t)]^{-d}$ which saturates at $[\ln L_{\rm sys}]^{-d}$ for 
$t\gtrsim \tau_c$. 

We also expect that, due to the frequency dependence 
of $\nu_{\rm eff}(\omega)$, the number of {\em diffusing} states depends on 
time as $N(1/t)$ with $N(\omega)$ defined in (\ref{eq:29}). 
The time dependent 
pair size therefore behaves as $L_1\,\ln(\Gamma_1 t)$ for 
$\Gamma_1^{-1}\ll t\lesssim \tau_c$ and as $L_1 \ln L_{\rm sys}$ for 
$\tau_c\lesssim t$. In recent numerical simulations \cite{shep3,shep4} 
of two ``interacting'' kicked rotators, Borgonovi and Shepelyansky indeed 
observed the logarithmic increase of the pair size with time. 

As a further illustration, we consider the probability that an electron 
pair performs a transition from a product state $|Rj>$ to another state 
$|\tilde R \tilde j>$ in the time $t>0$. This probability is given by
\begin{equation}
\label{eq:43}
\left|<\tilde R\tilde j|\,e^{-i\,{\cal H}t}\,|Rj>\right|^2=
\frac{1}{4\pi^2} \int dE\int d\omega\ e^{-i\omega t}\,
\left(G_{\tilde R R}^{\,\tilde j\,j}\right)^{(+)}(E+\frac{\omega}{2})\,
\left(G_{R\tilde R}^{\,j\,\tilde j}\right)^{(-)}(E-\frac{\omega}{2})\ ,
\end{equation}
where 
\begin{equation}
\label{eq:44}
\left(G_{\tilde R R}^{\,\tilde j\, j}\right)^{(\pm)}(E)
=<\tilde R\tilde j|\frac{1}{E\pm i\eps-{\cal H}}|Rj>
\end{equation}
denotes the matrix elements of the advanced or retarded Green's function. 
The ensemble average with respect to $\hat\zeta$ of the transition
probability can be calculated from the functional $F(J)$
(choosing a suitable source matrix 
$J$ and taking the derivatives). Omitting technical details, we give 
the result of a perturbative evaluation in the diffusive regime using 
the $\sigma$ model (\ref{eq:39}):
\begin{eqnarray}
\left\langle\left|<\tilde R\tilde j|\,e^{-i\,{\cal H}t}\,|Rj>\right|^2
\right\rangle_\zeta & \sim & \int dE\int d\omega\ e^{-i\omega t}
\nonumber\\
&&\nonumber\\
&&\times 
\left(\frac{\frac{1}{2\pi}\Gamma_1 v(\tilde j)}
{\left(E-\eta_{\tilde R}^{\tilde j}+\frac{1}{2}\Gamma_0 v(\tilde j)\right)^2
+\left(\frac{1}{2}\Gamma_1 v(\tilde j)-\frac{i}{2}\omega\right)^2}
\right)\ 
\nonumber\\
&&\times 
\left(\frac{\frac{1}{2\pi}\Gamma_1 v(j)}
{\left(E-\eta_{R}^{j}+\frac{1}{2}\Gamma_0 v(j)\right)^2
+\left(\frac{1}{2}\Gamma_1 v(j)-\frac{i}{2}\omega\right)^2}
\right)
\nonumber\\
&&\nonumber\\
&&\times (2\pi)^{-d}\int dq\ {\cal R}(q,\omega)\,e^{iq(R-\tilde R)}\quad.
\label{eq:45}
\end{eqnarray}
The contribution in the last line is the ``diffusion'' propagator in position 
and frequency space with ${\cal R}(q,\omega)=[\sigma_{\rm eff}\,q^2-i\omega
\,\nu_{\rm eff}(\omega)]^{-1}$ as above. Physically, it describes how the 
electron pair diffuses along its center of mass coordinate $R$ [with the 
modifications shown in (\ref{eq:42})]. The only information about the 
relative coordinate $j$ is contained in the Lorentzian prefactors
in (\ref{eq:42}). These effectively define a resonance condition 
for the diagonal energies: $|\eta_R^j-\eta_{\tilde R}^{\tilde j}|\lesssim 
\Gamma_1\,\mbox{min}[v(j),v(\tilde j)]$. For the well coupled product 
states ($v(j)\sim 1,\ v(\tilde j)\sim 1$), we recover that essentially states 
inside a band of width $\Gamma_1$ are coupled by the diffusive transport 
processes. For the badly coupled product states ($v(j)\sim e^{-4|j|/L_1},
\ v(\tilde j)\sim e^{-4|\tilde j|/L_1}$, with $|j|,|\tilde j|\gg L_1$) this 
resonance condition is practically never fulfilled and the transition 
probability becomes exponentially small due to the Lorentzian factors. 
However, for certain improbable realizations of the $\eta_R^j$ the resonance 
condition may be fulfilled even for two badly coupled product states. Then 
the corresponding transition probability experiences a strong enhancement.

\subsection {TIP Level Statistics at $L > L_1$}

The $\sigma$ model action (\ref{eq:39}) also provides a very convenient tool 
to discuss level correlations.
Actually, the two point correlation function, defined by
\begin{equation}
\label{eq:46}
Y_2(\omega)=\frac{1}{\langle \rho(E)\rangle\,\langle \rho(E+\omega)\rangle}
\Bigl(\langle \rho(E)\rangle\,\langle \rho(E+\omega)\rangle
-\langle \rho(E)\,\rho(E+\omega)\rangle\Bigr)\quad,
\end{equation}
($\rho(E)$ is the unaveraged total density of states)
can be calculated as in
Ref. \cite{efetov}. Of course, in this way, we only obtain the correlations 
of the well coupled (diffusing) pair states. The isolated pair states are 
mainly uncorrelated with the well coupled states and among themselves. 
Therefore, their contribution to the correlation function (\ref{eq:46}) 
essentially cancels out. For a finite system of size $L$, the diffusive 
dynamics defines an energy scale $E_c^{(2)}$, the ``pair Thouless 
energy'', which is in our case determined by the implicit equation 
${\cal D}_{\rm eff}(E_c^{(2)})/L^2=E_c^{(2)}$. For frequencies smaller 
than $E_c^{(2)}$ the second term of the action ${\cal L}_2[Q]$ in 
(\ref{eq:39}) dominates the level correlations. 
$Y_2(\omega)$ can 
exactly be expressed by the same integral (over one $Q$--matrix) as 
in \cite{efetov}. Therefore we recover the random matrix result
\begin{equation}
\label{eq:47}
Y_2(\omega)=Y_2^{(GOE)}\left(\frac{\omega}{\Delta(\omega)}\right)
\quad,\quad \Delta(\omega)^{-1}=L^d\,\nu_{\rm eff}(\omega)
\quad,\quad |\omega|\ll E_c^{(2)}
\end{equation}
with $\nu_{\rm eff}$ as in (\ref{eq:40}), (\ref{eq:41}). $\Delta(\omega)$ is a 
frequency dependent effective levelspacing and $Y_2^{(GOE)}(r)$ is the 
universal spectral correlation function of the Gaussian orthogonal ensemble 
($r$ is the energy difference measured in units of the level spacing). 
For higher frequencies, i.e. $E_c^{(2})
\ll|\omega|\ll \Gamma_1$ the first (kinetic) term of ${\cal L}_2[Q]$ has 
to be taken into account. The perturbative evaluation of $Y_2(\omega)$ is 
completely equivalent to the diagrammatical calculation of $Y_2(\omega)$ 
given by Alt'shuler and Shklovskii \cite{alt1}. We get
\begin{equation}
\label{eq:48}
Y_2(\omega)\sim \frac{1}{\omega^2}\,\left(\frac{\omega}
{{\cal D}_{\rm eff}(\omega)/L^2}\right)^{d/2}
\quad,\quad E_c^{(2)}\ll|\omega|\ll\Gamma_1\quad.
\end{equation}
Here ${\cal D}_{\rm eff}(\omega)/L^2$ can be viewed as a frequency dependent 
Thouless energy that sets the scale for the level correlations in the 
diffusive regime. 

In summary, we state that the field theoretical $\sigma$ model formulation 
of the effective random matrix model enabled us to recover many well 
known results for the localization problem of noninteracting particles.

\section{ Tightly Bound Particles in One Dimension} 

\label{section:3}

In this section, we are concerned with a particular example in one 
dimension which allows for a more microscopic treatment than the 
mapping onto the effective random matrix model. Our aim is to show 
that the physical implications and results found in the last section 
are not restricted to the random matrix approach and are indeed 
generic for the two interacting particle problem. Moreover, in the present
case we are able to generalize our treatment to more than two particles.

The essential idea is 
to compare the microscopic formulation of the problem of 
two interacting particles in a 1d random potential with the related problem
of independent particles in a quasi 1d disordered strip. To this end we start
with the case of two interacting particles and consider the Hamiltonian
\begin{equation}
H = -{1\over 2} (\partial_{x_1}^2 + \partial_{x_2}^2) + V(x_1) +V(x_2)
  + U(|x_1-x_2|) \quad , 
\label{eq1}
\end{equation}
where $x_1$ and $x_2$ are the coordinates of the two particles, 
$U(|x_1-x_2|)$ is the interaction potential, and $V(x)$ the 1d 
random potential with
\begin{eqnarray}
\langle V(x) \rangle &=& 0 \quad, \nonumber\\
\langle V(x) V(x') \rangle &=& c_1 \delta(x - x') \quad .
\label{eq2}
\end{eqnarray}
Introducing new coordinates $x=(x_1+x_2)/2$ and $y=x_1-x_2$ we have
\begin{equation}
H = -{1\over 4} \partial_x^2 + {1\over 2} H_T(y) + V(x+y/2) + V(x-y/2)
\label{eq3}
\end{equation}
with the transverse Hamiltonian
\begin{equation}
H_T(y) = -2\partial_y^2 + 2U(|y|) \quad .
\label{eq4}
\end{equation}
For later convenience we define a rescaled Hamiltonian $\tilde{H} = 2H$. 
We will
study the transport properties of $H$ at the two--particle energy 
$E$ (i.e., at $\tilde{E}=2E$ for $\tilde{H}$) and define a one--particle 
reference momentum $k_F$
by $E=2 k_F^2/2$. Next, we diagonalize the transverse Hamiltonian,
\begin{equation}
H_T(y) \phi_n(y) = (-2\partial_y^2+2U(|y|) \phi_n(y) = 
\epsilon_n\phi_n(y) \quad ,
\label{eq5}
\end{equation}
where the $\phi_n(y)$ are real eigenfunctions and we restrict ourselves 
to the even case, i.e. $\phi_n(y)=\phi_n(-y)$. Similar to Dorokhov 
\cite{dorokhov} we assume an
{\it attractive} potential between the two particles. In particular we consider
 a bag model, where $U(|y|)=\infty$ for $|y|> L$ and $U(|y|)=0$ otherwise. 
For this example the transverse eigenfunction system is simply given by
\begin{eqnarray}
\phi_n(y) &=& {1\over \sqrt{L}} \cos(k_n y) \quad , \nonumber \\
k_n &=& {\pi \over L} (n+{1\over 2}) \quad\quad (n=0,1,2,\ldots) \quad ,
    \nonumber \\
\epsilon_n &=& 2k_n^2 = 2\left( {\pi\over L}\right)^2 (n+{1\over 2})^2 \quad .
\label{eq8}
\end{eqnarray}
For the case of two electrons in a random potential an attractive 
interaction is of course
unrealistic. However, analytical arguments 
\cite{shep1,imry,fmgp} and in particular numerical studies 
based on the transfer matrix approach 
\cite{shep1,fmgpw} indicate that the sign of the interaction is 
basically immaterial for the transport properties of the system. In 
\cite{fmgpw} no essential
difference between a repulsive Hubbard--type interaction and an 
attractive bag interaction
was found. Therefore we feel justified to proceed with the
conceptually much simpler case of an attractive bag. Also, one might
think of a particle pair which is indeed (tightly) bound, e.g. an
exciton.

We project $\tilde{H}$ on the transverse eigenfunction system,
\begin{eqnarray}
\tilde{H}_{nm} &=& \langle \phi_n|\tilde{H}|\phi_m\rangle = 
\left[ -{1\over 2}\partial_x^2 + \epsilon_n \right] \delta_{nm} + 
\hat{V}_{nm}(x) \quad,
\nonumber\\
\hat{V}_{nm}(x) &=& 2\int_{-L}^L dy \ \phi_n(y) \phi_m(y) 
\left[ V(x+y/2) + V(x-y/2)\right] \quad .
\label{eq6}
\end{eqnarray}
Thus we have mapped the problem on a 1d, many channel Hamiltonian. 
We define a diagonal matrix
of longitudinal momenta,
\begin{equation}
\kappa = \mbox{diag} (\kappa_1\ldots\kappa_n\ldots); \quad\quad 
\kappa_n = \sqrt{2(\tilde{E}-\epsilon_n)} \quad .
\label{eq7}
\end{equation}
The $N$ open channels of the problem are defined by the condition 
$k_n = \sqrt{\epsilon_n/2} < k_F$.

\subsection{Mapping onto a Quasi-$1d$ Transfer Matrix Approach Without 
Interaction}

For comparison, let us now consider one electron in a disordered 2d strip,
\begin{equation}
H = -{1\over 2} (\partial_{x_1}^2 + \partial_{x_2}^2) + U(|x_2|) + V(x_1,x_2)
\quad .
\label{eq9}
\end{equation}
Here, $x_1$ and $x_2$ are the two space coordinates, $U(|x_2|)$ is the 
confining potential
defining the strip and $V(x_1,x_2)$ the 2d random potential with the 
statistical
properties
\begin{eqnarray}
\langle V(x_1,x_2) \rangle &=& 0 \quad , \nonumber \\
\langle V(x_1,x_2) V(x_1',x_2') \rangle &=& c_2 \delta(x_1-x_1') 
\delta(x_2-x_2')
\quad .
\label{eq10}
\end{eqnarray}
For notational simplicity we do not explicitly distinguish between the 
two cases considered here and hope that no confusion will arise. In close 
analogy to the previous problem of two particles we diagonalize the 
transverse Hamiltonian
\begin{equation}
H_T(x_2) = -{1\over 2}\partial_{x_2}^2 + U(|x_2|)
\label{eq11}
\end{equation}
so that
\begin{equation}
H_T(x_2) \phi_n(x_2) = \epsilon_n \phi_n(x_2) \quad .
\label{eq12}
\end{equation}
Again, we restrict ourselves to the even case $\phi_n(x_2)=\phi_n(-x_2)$. 
If we choose a box of width $2L$ with infinite walls as confining potential 
the transverse eigenfunction system is given by (\ref{eq8}). Projection of 
(\ref{eq9}) on the transverse modes gives
\begin{eqnarray}
H_{nm} &=& \langle \phi_n|H|\phi_m\rangle = 
\left[ -{1\over 2}\partial_{x_1}^2 + \epsilon_n \right] \delta_{nm} + 
\hat{V}_{nm}(x_1) \quad, \nonumber \\
\hat{V}_{nm} &=& \int_{-L}^L dx_2 \ \phi_n(x_2) \phi_m(x_2) V(x_1,x_2) \quad .
\label{eq13}
\end{eqnarray}
A matrix of longitudinal momenta can be defined as in (\ref{eq7}). The 
reference momentum $k_F$ of the two electron problem can now be interpreted 
as the one electron Fermi momentum and the open channels are characterized 
by the condition $k_n = \sqrt{2\epsilon_n} < k_F$. Comparing (\ref{eq6}) 
with (\ref{eq13}) we see that both the strip problem and the two electron 
problem are described by very similar quasi 1d (many channel) Hamiltonians. 
The only important difference resides in the definition of 
$\hat{V}_{nm}(x)$ in the two cases. In the next step we apply the 
transfer matrix method to (\ref{eq6}) and (\ref{eq13}).

We define a longitudinal $N$--channel wavefunction
$\psi(x)^T = (\psi_1(x), \ldots, \psi_N(x))$ so that the Schr\"odinger 
equation for a
Hamiltonian of the type (\ref{eq6}) or (\ref{eq13}) reads
\begin{equation}
E\psi(x) = H\psi(x) = -{1\over 2} \psi^{\prime\prime}(x) + \hat{\epsilon} 
\psi(x) + 
\hat{V}\psi(x) \quad ,
\label{eq14}
\end{equation}
where $\hat{\epsilon}$ is a diagonal matrix with $(\hat{\epsilon})_{nn} = 
\epsilon_n$ and
the primes in $\psi^{\prime\prime}(x)$ indicate the derivative with respect to 
$x$. We introduce a transfer matrix $A$ by
\begin{equation}
\left( \begin{array}{c}
       \psi(x) \\
       \psi'(x)
       \end{array}
\right)'
=
\left( \begin{array}{cc}
       0 & 1 \\
       2(\hat{\epsilon} - E + \hat{V} ) & 0 
       \end{array}
\right)
\left( \begin{array}{c}
       \psi(x) \\
       \psi'(x)
       \end {array}
\right)
\equiv
 A \left( \begin{array}{c}
       \psi(x) \\
       \psi'(x)
       \end {array}
\right) \quad .
\label{eq15}
\end{equation}
Redefining the wavefunction through
\begin{equation}
\tilde{v}(x) = {1\over\sqrt{2\kappa}}
\left( \begin{array}{c}
        \psi'(x) + i\kappa\psi(x) \\
        \psi'(x) - i\kappa\psi(x)
       \end{array}
\right)
\equiv
C \left( \begin{array}{c}
          \psi(x) \\
          \psi'(x)
         \end{array}
  \right)
\label{eq16}
\end{equation}
(we recall that $\kappa$ is the diagonal matrix (\ref{eq7})) we have
$\tilde{v}'(x) = CAC^{-1} \tilde{v}(x) \equiv \tilde{T}(x)\tilde{v}(x)$ with
\begin{equation}
\tilde{T}(x) =
i\left( \begin{array}{cc}
         \kappa & 0 \\
          0     & -\kappa
        \end{array}
  \right) -
{i\over\sqrt{\kappa}} \left( \begin{array}{cc}
                              \hat{V}(x) & -\hat{V}(x) \\
                              \hat{V}(x) & -\hat{V}(x)
                             \end{array}
                      \right)
{1\over\sqrt{\kappa}} \quad .
\label{eq17}
\end{equation}
To get rid of the first matrix on the r.h.s. of (\ref{eq17}) we redefine the
wavefunction again,
\begin{equation}
\tilde{v}(x) = \left( \begin{array}{cc}
                      e^{i\kappa x} & 0 \\
                      0 & e^{-i\kappa x}
                      \end{array}
                \right)
v(x) \quad ,
\label{eq18}
\end{equation}
and finally arrive at
\begin{equation}
v'(x) = T(x) v(x)
\label{eq19}
\end{equation}
with
\begin{equation}
T(x) = -{i\over \sqrt{\kappa}}
\left( \begin{array}{cc}
   e^{-i\kappa x} \hat{V}(x) e^{i\kappa x} & -e^{-i\kappa x} \hat{V}(x) 
e^{-i\kappa x} \\
   e^{ i\kappa x} \hat{V}(x) e^{i\kappa x} & -e^{ i\kappa x} \hat{V}(x) 
e^{-i\kappa x} 
       \end{array}
\right)
{1\over \sqrt{\kappa}} \quad .
\label{eq20}
\end{equation}
It is not difficult to see that (\ref{eq19}) is equivalent to the 
following evolution law
for the full transfer matrix $\hat{M}(x)$,
\begin{equation}
\hat{M}(x+\delta x) = e^{\delta \hat{M}} \hat{M}(x),
\label{eq21}
\end{equation}
where the ``building block'' $\delta \hat{M}$ is given by
\begin{equation}
\delta \hat{M} = \int_x^{x+\delta x} T(x')dx' = 
-{i\over\sqrt{\kappa}} 
\left( \begin{array}{cc}
        \delta v_1 & -\delta v_2 \\
        \delta v_2^\dagger & -\delta v_1^\dagger
       \end{array}
\right)
{1\over \sqrt{\kappa}}
\equiv
\left( \begin{array}{cc}
        a & b \\
        b^\dagger & d
       \end{array}
\right) 
\label{eq22}
\end{equation}
with
\begin{eqnarray}
\delta v_1 &=& \int_x^{x+\delta x} dx' e^{-i\kappa x'} \hat{V}(x') 
e^{i\kappa x'}
    \quad ,          \nonumber \\
\delta v_2 &=& \int_x^{x+\delta x} dx' e^{-i\kappa x'} \hat{V}(x') 
e^{-i\kappa x'}
\quad .
\label{eq23}
\end{eqnarray}
Under the crucial assumptions that both the blocks of $T(x)$ in
(\ref{eq20}) and all the matrix elements within these blocks are
statistically independent, we deal with the so--called isotropic
situation and the evolution law (\ref{eq21}) leads to the
Dorokhov--Mello--Pereyra--Kumar (DMPK) equation \cite{fokpla} for the
eigenvalue distribution of $\hat{M}(x)$. This equation accurately describes
the transport of independent electrons in quasi 1d wires and is --- at
least in the unitary case --- completely under control 
\cite{GUEcase,frahm_let}.
In the case of a 2d strip of width not significantly exceeding the
elastic mean free path $l_{el}$ the exponential factors $e^{\pm
  i\kappa x}$ in (\ref{eq20}) oscillate rapidly so that the assumption
of statistically independent matrix elements seems very reasonable.
This argument provides a microscopic justification for the application
of the DMPK equation to quasi 1d wires \cite{fokpla}.  A main purpose
of our present study is to point out that the DMPK equation also
describes the transport of two interacting electrons in a disordered
1d system, provided the pair size does not exceed $l_{el}$. This
statement should be practically obvious since both the strip and the
two electron problem have been formulated in almost identical
technical terms: The DMPK equation is as well--justified for the
latter problem as for the former.

\subsection{ TIP-Localization Length}

Once the applicability of DMPK theory is guaranteed the next step is
to identify the localization length of the problem in terms of the
given system parameters. The connection between the building block
(\ref{eq22}) and the localization length $\xi$ in the DMPK approach is
given by \cite{fokpla}
\begin{equation}
{N^2\over \xi} \delta x = \langle tr(b^\dagger b) \rangle \quad ,
\label{eq24}
\end{equation}
i.e.
\begin{equation}
{N^2\over \xi} \delta x = \int_x^{x+\delta x} dx_1 dx_2 
\left\langle tr \left[ {1\over\kappa} e^{-i\kappa(x_1-x_2)} 
                        \hat{V}(x_1)  
                       {1\over\kappa} e^{-i\kappa(x_1-x_2)}
                        \hat{V}(x_2) 
                \right] \right\rangle \quad .
\label{eq25}
\end{equation}
Any further evaluation of (\ref{eq25}) depends on the precise form of 
$\hat{V}(x)$ and there, finally, the differences between the two cases
studied here come in.

Let us begin with the disordered strip. Inserting $\hat{V}(x)$ as defined
in (\ref{eq13}) in (\ref{eq25}) we get
\begin{eqnarray}
{N^2\over \xi} \delta x &=& \int_x^{x+\delta x} dx_1 dx_2 
                          \int_{-L}^L dy_1 dy_2
 tr \left[ {1\over\kappa} e^{-i\kappa(x_1-x_2)} 
                     \phi(y_1)\phi(y_1)^T
                       {1\over\kappa} e^{-i\kappa(x_1-x_2)}
                     \phi(y_2)\phi(y_2)^T 
                \right]  \nonumber\\
& & \quad \langle V(x_1,y_1) V(x_2,y_2) \rangle
\label{eq26}
\end{eqnarray}
and after using (\ref{eq10}) and then (\ref{eq8})
\begin{eqnarray}
{N^2\over \xi} \delta x &=& c_2 \delta x \int_{-L}^L dy
          \sum_{nm} {1\over \kappa_n\kappa_m}
          \phi_n(y)^2 \phi_m(y)^2 \nonumber\\
 &=& {c_2\over 2L} \delta x \sum_{nm} {1\over \kappa_n\kappa_m}
     \left( 1+{1\over 2} \delta_{nm}\right) \quad .
\label{eq27}
\end{eqnarray}
Evaluating the remaining sums approximately 
(see appendix \ref{app:axel_a}),
\begin{eqnarray}
\sum_{n=0}^{N-1} {1\over\kappa_n} &\approx& {L\over 2} \quad,
\nonumber\\
\sum_{n=0}^{N-1} {1\over \kappa_n^2} &\approx& {L^2\over 2\pi^2} 
{\ln 2N\over N}
\label{eq28}
\end{eqnarray}
we can neglect the diagonal contribution in (\ref{eq27}) as a 
$1/N$ correction. 
Together with (see appendix \ref{app:axel_b})
\begin{equation}
c_d \approx {\pi^{d-1}\over d} {k_F^{4-d}\over k_F l_{el}}
\label{eq29}
\end{equation}
for $d=2$ we finally arrive at
\begin{equation}
{N^2\over \xi} = {\pi\over 16} {k_FL\over l_{el}}
\Leftrightarrow
\xi = {16\over\pi^2} Nl_{el} \quad.
\label{eq30}
\end{equation}
This is the well--known relation between the localization length and the
elastic mean free path in quasi 1d disordered strips. In the following
we will show that a similar relation also holds in the case of two electrons
with a bag interaction, although the detailed reasoning is a little more
involved than in the previous case. From (\ref{eq6}) and (\ref{eq25}) we have
(instead of (\ref{eq26}))
\begin{eqnarray}
{N^2\over \xi} \delta x &=& 4 \int_x^{x+\delta x} dx_1 dx_2 
                          \int_{-L}^L dy_1 dy_2
 tr \left[ {1\over\kappa} e^{-i\kappa(x_1-x_2)} 
                     \phi(y_1)\phi(y_1)^T
                       {1\over\kappa} e^{-i\kappa(x_1-x_2)}
                     \phi(y_2)\phi(y_2)^T 
                \right]  \nonumber\\
& & \quad 
\langle
\left[ V(x_1+y_1/2) + V(x_1-y_1/2) \right]  
\left[ V(x_2+y_2/2) + V(x_2-y_2/2) \right]
\rangle
\quad .
\label{eq31}
\end{eqnarray}
According to (\ref{eq2}) the average over the random potential leads to
four $\delta$--functions so that $x_1-x_2 = (s_1y_1+s_2y_2)/2$ with
$s_{1,2} = \pm 1$ (independently). Since the typical scale for $\delta x$ is
$l_{el}$ the $\delta$--function condition can be always fulfilled and we 
arrive at
\begin{eqnarray}
{N^2\over \xi} \delta x &=&
4c_1\delta x \sum_{s_1,s_2=\pm 1} tr \left[
\left( {1\over\kappa}\int dy_1 \ e^{-i\kappa s_1y_1/2} \phi(y_1)\phi(y_1)^T
                             e^{-i\kappa s_1y_1/2} \right) \right. \nonumber\\
& &
\phantom{4c_1\delta x \sum_{s_1,s_2=\pm 1} tr [}
\left. \left( {1\over\kappa}\int dy_2 \ e^{-i\kappa s_2y_2/2} 
\phi(y_2)\phi(y_2)^T
                             e^{-i\kappa s_2y_2/2} \right) \right]
\label{eq32}
\end{eqnarray}
or
\begin{equation}
{N^2 \over \xi} = 16 c_1 \ tr\left[ {1\over\kappa} B {1\over\kappa} B \right]
\quad ,
\label{eq33}
\end{equation}
where
\begin{equation}
B_{nm} = \int_{-L}^{L} dy \cos((\kappa_n+\kappa_m)y/2) \phi_n(y) \phi_m(y)
\quad .
\label{eq34}
\end{equation}
Employing (\ref{eq29}) for $d=1$ and introducing 
$\eta_n=\sqrt{N^2-(n+1/2)^2}$, (\ref{eq33}) can be rewritten as
\begin{equation}
{1\over \xi} = {4\over l_{el}} \sum_{nm} {1\over \eta_n\eta_m} B_{nm}^2
\quad .
\label{eq35}
\end{equation}
In appendix \ref{app:axel_c} it is shown that (\ref{eq35}) leads 
to the final result
\begin{equation}
\xi \approx {N l_{el} \over 2\sqrt{2} \ln(1+\sqrt{2}) } \quad .
\label{eq36}
\end{equation}
Using $N=k_FL/\pi$ and $L\approx l_{el}$ (which is the maximal pair size
allowed within the present technical framework) we have
\begin{equation}
\xi \sim k_Fl_{el}^2 \quad ,
\label{eq37}
\end{equation}
a result which has first been derived by Dorokhov 
\cite{dorokhov} for two particles
interacting via a harmonic potential. Eq. (\ref{eq37}) shows that the 
localization length for two interacting particles is significantly
enhanced over the one--particle value $\xi\sim l_{el}$.

\subsection{Scaling of the TIP-Conductance}

In conclusion, we have demonstrated the validity of the DMPK equation for
the transport properties of two electrons in a narrow bag in 1d. 
As our main argument we have pointed out the almost complete analogy to the
the case of a disordered quasi 1d strip. Identifying the localization length
in terms of the system parameters we recovered Dorokhov's result (\ref{eq37}).

As probably the most important consequence of our investigation we can
identify, as in the $\sigma$ model formulation above,
a {\it scaling parameter} in the two electron problem, namely
the conductance 
$g=2 \, {\rm tr}[(\hat{M}^\dagger \hat{M} + (\hat{M}^\dagger \hat{M})^{-1} + 2)^{-1}]$ 
of the associated strip problem.
Although this quantity does not seem to have an immediate physical
interpretation for two interacting electrons and should perhaps be
regarded as some sort of auxiliary variable it is nevertheless of
considerable value: Its length dependence in the whole range from the
diffusive to the localized regime is under control \cite{mmz} and the
crossover point between these regimes is given by $g(\xi)=1$. In this
way we have again justified the assumptions that underly Imry's
application \cite{imry} of the Thouless block picture to the problem
of interacting particles in a random potential. We think that Imry's
pair conductance $g_2$ should be identified with our conductance $g$
of the associated strip problem. This identification explains the
remarkable success of the scaling block picture in the present context
and clarifies the nature of the scaling parameter.

\subsection{Generalization to Many Particles}

 The above considerations for two interacting particles in a random
potential can be generalized to larger particle numbers provided the
particle density (in the case of Fermions) is low enough to neglect
effects of the Pauli principle. In this paper we refrain from discussing
the full many body problem, where quasiparticles are the relevant degrees
of freedom. A particularly important issue in this context would be the
lifetime of interacting quasiparticles as a function of their excitation energy.
For a first numerical study of the
influence of the Pauli principle on interaction--assisted coherent
transport see \cite{oppen2}.

Let us consider the $M$--particle Hamiltonian
\begin{equation}
H = -{1\over 2} \sum_{i=1}^M \left( \partial_{x_i}^2 + V(x_i) \right)
    + \sum_{i<j} U(|x_i-x_j|)   \quad,
\label{eq2.1}
\end{equation}
where $U(|x-x'|)$, as before, represents an attractive bag of size
$2L$. Introducing the center of mass coordinate $x$ and $M-1$ relative
coordinates,
\begin{eqnarray}
x &=& {1\over M} \sum_{i=1}^M x_i  \quad,      \nonumber\\
y_j &=& x_j-x_{j+1} \quad\quad (j=1,\ldots,M-1) \quad,
\label{eq2.2}
\end{eqnarray}
we get
\begin{eqnarray}
H &=& -{1\over 2M} \partial_x^2 -
f(\partial_{y_1},\ldots,\partial_{y_{M-1}})
+ \sum_{i=1}^M V(x+g_i(y_1,\ldots,y_{M-1}))  \nonumber\\
& & \ + \sum_{i<j} U(h_{ij}(y_1,\ldots,y_{M-1})) \quad.
\label{eq2.3}
\end{eqnarray}
Here, $f$, $g_i$, and $h_{ij}$ are functions which we do not  have to
specify in detail for our subsequent considerations. In complete
analogy to the previous section we can define a transverse Hamiltonian
and its eigenfunction system by
\begin{eqnarray}
& & H_T = -f(\partial_{y_1},\ldots,\partial_{y_{M-1}}) 
 + \sum_{i<j} U(h_{ij}(y_1,\ldots,y_{M-1})) \quad, \nonumber\\
& & H_T \varphi_{n_1\ldots n_{M-1}}(y_1,\ldots,y_{M-1}) = 
\epsilon_{n_1\ldots n_{M-1}} \varphi_{n_1\ldots
  n_{M-1}}(y_1,\ldots,y_{M-1}) \quad.
\label{eq2.4}
\end{eqnarray}
We do not specify any particular symmetry for the transverse
eigenfunctions
$\varphi_{n_1\ldots n_{M-1}}(y_1,\ldots,y_{M-1})$.
Projecting the full Hamiltonian $H$ on the transverse eigenfunction
system we arrive at
\begin{equation}
H_{n_1'\ldots n_{M-1}'}^{n_1\ldots n_{M-1}} = 
\left( -{1\over 2M}\partial_x^2 + \epsilon_{n_1\ldots n_{M-1}} \right)
\delta_{n_1n_1'} \ldots \delta_{n_{M-1}n_{M-1}'} + 
\hat{V}_{n_1'\ldots n_{M-1}'}^{n_1\ldots n_{M-1}}(x) \quad,
\label{eq2.5}
\end{equation}
where
\begin{eqnarray}
& &\hat{V}_{n_1'\ldots n_{M-1}'}^{n_1\ldots n_{M-1}}(x) =
\int dy_1\ldots dy_{M-1} \ 
\varphi_{n_1\ldots n_{M-1}}(y_1,\ldots,y_{M-1}) \nonumber\\
& & \hspace{1cm}\left(
\sum_{i=1}^M V(x+g_i(y_1,\ldots,y_{M-1})) 
\right)
\varphi_{n_1'\ldots n_{M-1}'}(y_1,\ldots,y_{M-1}) \quad.
\label{eq2.6}
\end{eqnarray}
Defining the ``building block'' $\delta M$ as in (\ref{eq22}) and
keeping in mind the relation (\ref{eq24}) we have
\begin{equation}
\langle {\rm tr} (b^\dagger b) \rangle = 
\int_x^{x+\delta x} dx_1dx_2 \ 
\left\langle {\rm tr}
\left( 
{1\over \kappa} e^{-i\kappa(x_1-x_2)} \hat{V}(x_1) {1\over \kappa}
                e^{-i\kappa(x_1-x_2)} \hat{V}(x_2) 
\right)
\right\rangle \quad,
\label{eq2.7}
\end{equation}
where $\kappa_{n_1\ldots n_{M-1}} = \sqrt{2(E-\epsilon_{n_1\ldots
    n_{M-1}})}$. The relevant average in (\ref{eq2.7}) is
straightforwardly calculated,
\begin{eqnarray}
& &\left\langle
\sum_{i=1}^M V(x_1+g_i(y_1 \ldots y_{M-1} ))
\sum_{j=1}^M V(x_2+g_j(y_1'\ldots y_{M-1}'))
\right\rangle =                                  \nonumber\\
& & \hspace{2cm}
{k_F^3\over 2k_fl_{el}}
\sum_{i,j=1}^M \delta(x_1-x_2+g_i(y_1 \ldots y_{M-1} )-
                              g_j(y_1'\ldots y_{M-1}')) \quad.
\label{eq2.8}
\end{eqnarray}
Both the $x_1$  and the $x_2$ integration in (\ref{eq2.7}) can be
easily performed and we get
\begin{eqnarray}
\langle {\rm tr}(b^\dagger b) \rangle &=&
{k_F^3\over 2k_Fl_{el}} \delta x 
\sum_{i,j=1}^M \int dy_1\ldots dy_{M-1} dy_1' \ldots dy_{M-1}'
\nonumber\\
& & \hspace{-1cm}
{\rm tr}
\left(
{1\over\kappa} e^{-i\kappa(g_j-g_i)} 
\varphi(y_1,\ldots,y_{M-1})\varphi(y_1,\ldots,y_{M-1})^T 
\right. \nonumber\\
& & 
\left. {1\over\kappa} e^{-i\kappa(g_j-g_i)} 
\varphi(y_1',\ldots,y_{M-1}')\varphi(y_1',\ldots,y_{M-1}')^T
\right)                        \nonumber\\
&=& 
{k_F^3\over 2k_Fl_{el}} \delta x
{\rm tr} 
\left( {1\over\kappa} B {1\over\kappa} B^\dagger  \right)
\label{eq2.9}
\end{eqnarray}
with
\begin{eqnarray}
B^{n_1\ldots n_{M-1}}_{n_1'\ldots n_{M-1}'} &=&
\int_{-L}^L dy_1 \ldots dy_{M-1}
\left(
\sum_{i=1}^M e^{i(\kappa_{n_1\ldots n_{M-1}} + \kappa_{n_1'\ldots
    n_{M-1}'}) g_i(y_1,\ldots,y_{M-1})}
\right) \nonumber\\
& & \quad
\varphi_{n_1\ldots n_{M-1}}(y_1,\ldots,y_{M-1})
\varphi_{n_1'\ldots n_{M-1}'}(y_1,\ldots,y_{M-1})  \quad,
\label{eq2.10}
\end{eqnarray}
which corresponds to (\ref{eq34}). For further progress we have to
rely on certain estimates. Let us assume that the phase factor
$\exp(i(\kappa_{n_1\ldots n_{M-1}} + \kappa_{n_1'\ldots
    n_{M-1}'}) g_i(y_1,\ldots,y_{M-1}))$ fluctuates strongly when the
$y_i$ vary between $-L$ and $L$. Then it can be replaced by its
average denoted by $c_{ph}$ and we have
\begin{equation}
B^{n_1\ldots n_{M-1}}_{n_1'\ldots n_{M-1}'} \approx
c_{ph} \delta_{n_1\ldots n_1'} \ldots \delta_{n_{M-1}n_{M-1}'} \quad.
\label{eq2.11}
\end{equation}
Our assumption is only correct if $\kappa_{n_1\ldots n_{M-1}} \gg
1/L$. This means that channels with high transverse excitation
energies and small longitudinal momenta $\kappa_{n_1\ldots n_{M-1}}$
are not correctly described by our approximation. Comparing the result
of our estimate with the more accurate calculation for $M=2$ in the
previous section we can determine the error incurred. Using
(\ref{eq24}) and (\ref{eq2.11}) we have
\begin{equation}
{1\over\xi} = {c_{ph}^2 \over N^2} {k_F^3\over 2k_Fl_{el}}
\sum_{n_1\ldots n_{M-1}} {1\over \kappa^2_{n_1\ldots n_{M-1}}}
\quad.
\label{eq2.12}
\end{equation}
We estimate the sum in (\ref{eq2.12}) by an integral approximation,
\begin{equation}
\sum_{n_1\ldots n_{M-1}} {1\over \kappa^2_{n_1\ldots n_{M-1}}}
\approx
\int_0^E d\epsilon \ \rho(\epsilon) {1\over 2(E-\epsilon)}
\quad,
\label{eq2.13}
\end{equation}
where $E$ is the $M$--particle energy at which we study the transport
problem, and $\rho(\epsilon)$ is the density of transverse energy
levels. To continue, we replace $\rho(\epsilon)$ by its average
$\bar{\rho}$ and assume that $E-\epsilon \approx \Delta$ (with
$\Delta$ the transverse level spacing) provides the cutoff for the
otherwise divergent integral in (\ref{eq2.13}):
\begin{eqnarray}
\int_0^E d\epsilon\ \rho(\epsilon) {1\over 2(E-\epsilon)}
&=& {1\over 2} \int_0^1 dx {\rho(x)\over 1-x} \nonumber\\
&\approx& {\bar{\rho}\over 2} \int_0^{1-\delta} {dx\over 1-x}
\nonumber\\
&=& -{\bar{\rho}\over 2} \ln\delta \quad.
\label{eq2.14}
\end{eqnarray}
With $E\sim k_F^2$, the number of transverse channels
$N=(k_FL)^{M-1}$, $\bar{\rho}=N/E$, and $\delta=\Delta/E=1/N$ we
finally arrive at (disregarding numerical prefactors)
\begin{equation}
\xi \sim {1\over c_{ph}^2} {(k_FL)^{M-1} l_{el} \over
  \ln[(k_FL)^{M-1}]} \quad.
\label{eq2.15}
\end{equation}
For the case $M=2$ we almost recover the result (\ref{eq37}) of the
previous section. The logarithmic suppression in (\ref{eq2.15}) is an
artifact of our approximations. It arises precisely from the levels
with $E\approx\epsilon$, i.e. $\kappa\approx 0$, which are not
well--described by our approach. Therefore we discard the
$\ln$--correction in (\ref{eq2.15}) and have as our final result (for
$L\approx l_{el}$)
\begin{equation}
\xi \sim (k_Fl_{el})^{M-1} l_{el} \quad.
\label{eq2.16}
\end{equation}
This shows how the enhancement factor for coherent multiple particle
propagation increases with particle number $M$.

\section{Conclusion and Discussion}

In this work, we have developed and further improved two analytical 
approaches to the phenomenon of
interaction--assisted coherent pair propagation. One, the $\sigma$ model formulation 
based on an effective random Hamiltonian, was inspired by Shepelyansky's view \cite{shep1}.
The other, a transfer matrix approach to two or many tightly bound particles, was built on
the pioneering ideas of Dorokhov \cite{dorokhov}. In its most succinct form, the message from
our investigations is the following: {\it Transport of interacting pairs or of small aggregates
of particles can be described in terms which have been introduced in the context of, and are 
therefore familiar from, one--electron theory}.

In particular, pair transport is governed by a scaling parameter, the effective two--particle
conductivity. This is a rigorous statement, once the models introduced above are 
accepted, and adds to our confidence in Imry's generalization \cite{imry} of the Thouless
scaling picture. Furthermore, there is an energy scale, the two--particle Thouless energy,
associated with diffusive pair transport. It corresponds, as in one--electron theory, to
the time it takes to diffuse through the sample and defines the scale, for which deviations from
random matrix theory occur in the spectral statistic of diffusing pair states. The precise pair
dynamics, however, differs from the one--particle case in that both the diffusion constant and the
density of states are frequency dependent. This is a clear manifestation of the pairs' composite
character, since it is connected with a logarithmic growth of the pair size and a corresponding 
logarithmic attenuation of the diffusion constant as a function of time. Apart from confirming
important aspects of these conclusions, our second
(transfer matrix) approach enabled us to study, in a
rather general form, the behavior of more than two tightly bound particles. It turned out that the
enhancement factor, i.e. the ratio between the $M$--particle and the one--particle localization
length, grows as the $(M-1)$th power of the latter.

Of course, both models introduced in this paper have their shortcomings. Foremost, our $\sigma$
model formulation relies heavily on the assumption that the effective Hamiltonian catches the
essential physics of the problem. We can imagine two types of criticism: (i) {\it quantitative}
statistical assumptions like the size of the fluctuating interaction matrix elements or the
form of the decay towards the band edges are incorrect or unrealistic. Corresponding modifications 
would lead to quantitative corrections without precluding our whole approach. (ii) {\it Qualitative}
assumptions like the statistical independence of the matrix elements are incorrect in an essential
way. This would call into question the $\sigma$ model formulation as such, since it 
is hard to imagine how a similar $\sigma$ model can be obtained in the presence of important
correlations. One should also note that the one--particle localization length $L_1$
is the {\it smallest} length scale resolved by the effective Hamiltonian. For length scales
$L<L_1$ effects associated with the transport of single particles have to be taken into account
\cite{akpi}. In this case the interpretation of the two--particle conductance is less
straightforward.

The major drawback of the transfer matrix model, on the other hand,
is its restriction to tightly bound particles. This restriction
is necessary to prevent the disorder from being
relevant for the {\it relative} motion of the particles. Otherwise relative and center of mass
motion would have been coupled and the problem much harder to solve. Naturally, however, the
requirement of strong binding restricts the physical situations to which the model may be applied.

Combined, the two approaches explained in this paper represent our analytical knowledge concerning
interaction--assissted coherent transport. For the future, we believe that it is an important and
worthwhile task to devise and investigate an effective Hamiltonian for more than two particles.
This would be another step towards understanding many--particle effects and it would be free of
the constraints inevitably associated with the transfer matrix method. It is also of interest to
investigate the question, whether the effect discussed here is related to a mechanism for the
enhancement of persistent currents proposed some time ago \cite{pc}.

\section*{Acknowledgment}

We are grateful to Y. Imry, D.L. Shepelyansky and D. Weinmann for fruitful discussions and their
interest in our work. AMG would also like to acknowledge discussions with S. Grossmann,
D. Lohse, A.D. Mirlin, F. von Oppen, R. R\"omer, M. Schreiber, and P. W\"olfle. This work was
supported by the European HCM program (KF) and a NATO fellowship through the DAAD (AMG).

\appendix

\section{Expansion of the effective potential (\ref{eq:32})}

\label{app:klaus_a}

In this Appendix, we derive an expression for the effective $Q$--potential 
$f_R(Q)$ which is more suitable for the comparison with the standard 
$\sigma$ model of a disordered metal ${\cal L}_{met}[Q]$. For this, we 
consider a quantity like 
\begin{equation}
\label{eq:a1}
I=\mbox{str}\ \ln(a+b\Lambda+c Q)
\end{equation}
with some complex parameters $a,\ b,\ c$ and $\Lambda$, $Q$ as in 
(\ref{eq:32}). In the grading for advanced and retarded Green's functions, 
$\Lambda$ takes the form $\Lambda={1\ \ 0\choose 0\ -1}$. The quantity $I$ 
only depends on the radial parameters $\theta$ \cite{efetov} of 
$Q=T^{-1}\Lambda T$, i.e. we may choose $T=\exp(\theta\sigma_1)$, where 
$\sigma_1={0\ 1\choose 1\ 0}$ and $\theta$ is $4\times 4$ supermatrix with 
two or three independent radial parameters \cite{efetov} (depending on the 
symmetry class). Then we have $\sigma_1\Lambda\sigma_1=-\Lambda$ and 
$\sigma_1 Q\sigma_1=-Q$. Inserting $1=\sigma_1^2$ and permuting the matrix 
products in the argument of the logarithm in (\ref{eq:a1}), we thus obtain:
\begin{eqnarray}
I & = & \mbox{str}\ \ln(a-b\Lambda-c Q)=
\frac{1}{2}\mbox{str}\ \ln\Bigl((a+b\Lambda+c Q)(a-b\Lambda-c Q)\Bigr)=
\nonumber\\
& = & \frac{1}{2}\mbox{str}\ \ln\biggl(\Bigl[a^2-(b+c)^2\Bigr]-bc
\Bigl[Q\Lambda+\Lambda Q-2\Bigr]\biggr)\quad.
\label{eq:a2}
\end{eqnarray}
We now expand $I$, for the case in which either $|bc|\ll |a^2-(b+c)^2|$ is 
valid or the matrix $\Delta Q=(Q\Lambda+\Lambda Q-2)$ may be considered as 
small. The latter condition is clearly valid for a perturbative treatment 
where $Q$ can be parameterized \cite{efetov} as $Q=\Lambda\frac{1+iP}{1-iP}
\simeq \Lambda(1+2iP+(2iP)^2+\cdots)$ with $\Lambda P\Lambda=-P$. Then 
$\Delta Q=8(iP)^2+{\cal O}(P^4)$ is indeed a small parameter and expanding 
(\ref{eq:a2}), we find:
\begin{equation}
\label{eq:a3}
I\simeq -\frac{bc}{a^2-(b+c)^2}\ \mbox{str}(Q\Lambda)\quad.
\end{equation}
Applying (\ref{eq:a3}) on $f_R(Q)$ defined in (\ref{eq:32}), we obtain:
\begin{equation}
\label{eq:a4}
f_R(Q)\simeq -\frac{1}{2}\left(
\sum_j \frac
{(\frac{1}{2}\omega)(\frac{i}{2}\Gamma_1\,v(j))}
{(E-\eta_j^R+\frac{1}{2}\Gamma_0 v(j))^2-(\frac{1}{2}\omega+\frac{i}{2}\Gamma_1
\,v(j))^2}\right)\ \mbox{str}(Q\Lambda)\quad.
\end{equation}
Here, the terms with $|j|\lesssim L_c\sim L_1\ln L_{\rm sys}$ 
can be considered as 
selfaveraging with respect to the random variables $\eta_j^R$, due to 
overlapping resonances. The typical (most probable) values of the remaing 
terms (with $|j|\gtrsim L_c$) are exponentially small since $v(j)\sim 
e^{-4|j|/L_1}$. They can be neglected whereas the other terms may be 
replaced by their $\eta$--average. In addition, we assume that the 
integration range $2W_b$ for $\eta$ is much larger than the other energy 
scales $\omega$, $\Gamma_0$, $\Gamma_1$ and that the $\eta$--density 
$\rho_0(\eta)$ varies rather slowly on these ``small'' scales. The 
$\eta$--integral may then be extended to all $\eta$ with 
$-\infty<\eta<\infty$ and $\rho_0(\eta)$ can be replaced by $\rho_0(E)$. 
Performing the $\eta$--integration, we find:
\begin{equation}
\label{eq:a5}
f_R(Q)\simeq -i\frac{\pi}{4}\omega\ h\left(\frac{\Gamma_1}{\omega}\right)
\,\rho_0(E)\ \mbox{str}(Q\Lambda)
\end{equation}
with
\begin{equation}
\label{eq:a6}
h \left(\frac{\Gamma_1}{\omega}\right)=i\frac{\Gamma_1}{\omega}
\sum_{|j|\lesssim L_c}\frac{v(j)}{1+i\frac{\Gamma_1}{\omega} v(j)}
\quad.
\end{equation}
In order to obtain a more explicit expression for this function, we would 
have to specify the choice for $v(j)$. However, the relevant limiting cases 
are:
\begin{eqnarray}
h\left(\frac{\Gamma_1}{\omega}\right)& \simeq &
i\frac{\Gamma_1}{\omega}\left(S_1-i\frac{\Gamma_1}{\omega} S_2+\cdots
\right)\quad,\quad S_\nu=\sum_j [v(j)]^\nu\quad,\quad
\Gamma_1\ll |\omega|\quad,
\label{eq:a7}\\
&&\nonumber\\
h\left(\frac{\Gamma_1}{\omega}\right)& \simeq &
\left[\frac{L_1}{4}\ln\left(\frac{\Gamma_1}{|\omega|}\right)\right]^d
\quad,\quad \tau_c^{-1}\lesssim |\omega|\ll \Gamma_1\quad,
\label{eq:a8}\\
&&\nonumber\\
h\left(\frac{\Gamma_1}{\omega}\right)& \simeq &
\left[\frac{L_1}{4}\ln\left(\Gamma_1\tau_c\right)\right]^d=L_c^d
\quad,\quad  |\omega|\ll \tau_c^{-1}\quad.
\label{eq:a9}
\end{eqnarray}
Here, $\tau_c^{-1}=\Delta_{\rm eff}=\Gamma_1 v(L_c)$ is the effective 
level spacing of the well coupled pair states, introduced subsequent to 
Eq. (\ref{eq:28}). The cutoff in the sum (\ref{eq:a6}) is either given by 
$|j|\lesssim L_{\rm eff}(\omega)$ [cp. (\ref{eq:29})] or $|j|\lesssim L_c$, 
resulting in (\ref{eq:a8}) and (\ref{eq:a9}), respectively.

\section{Some Estimates}

\label{app:axel_a}

In this appendix we estimate the discrete sums $\sum_n (1/\kappa_n)$,
$\sum_n (1/\kappa_n^2)$ contributing in (\ref{eq27}). With
$2E\approx N^2\pi^2/L^2$,
$k_n^2 = \pi^2(n+1/2)^2/L^2\approx n^2\pi^2/L^2$,
and approximating the sum by an integral we have
\begin{eqnarray}
\sum_{n=0}^{N-1} {1\over \kappa_n} &\approx& \int_0^N {dn\over\sqrt{2E-k_n^2}} 
\nonumber\\
&\approx& {L\over \pi N} \int_0^N {dn\over \sqrt{1-n^2/N^2}}
\nonumber\\
&=& {L\over\pi} \int_0^1 {dx\over\sqrt{1-x^2}}
\nonumber\\
&=& L/2 \quad.
\label{eqa1}
\end{eqnarray}
Analogously, we get
\begin{equation}
\sum_{n=0}^{N-1} {1\over \kappa_n^2} \approx
{L^2\over \pi^2N} \int_0^1 {dx\over 1-x^2} \quad ,
\label{eqa2}
\end{equation}
which is logarithmically divergent. However, we have to take into account
the energy difference between $E$ and the transverse energy of the last 
channel.
This energy gap is of the order of the level spacing and hence of relative
magnitude $1/N$. Therefore,
\begin{eqnarray}
\sum_{n=0}^{N-1} {1\over\kappa_n^2} &\approx& {L^2\over \pi^2N}
\int_0^{1-1/N} {dx\over 1-x^2}   \nonumber\\
&=& {L^2\over \pi^2 N} {1\over 2} \ln{1+x\over 1-x} \Big\vert^{1-1/N}_0
\nonumber\\
&\approx& {L^2\over 2\pi^2} {\ln 2N \over N} \quad .
\label{eqa3}
\end{eqnarray}
This completes the derivation of (\ref{eq28}).

\section{Normalization of the Random Potential}

\label{app:axel_b}

To estimate the constants $c_1$ and $c_2$ appearing in (\ref{eq2}) and
(\ref{eq10}), respectively, we recall the normalization condition for models
with random white noise disorder \cite{efetov},
\begin{equation}
\langle V(\vec{x}) V(\vec{x}^{\ \prime}) = 
{1\over 2\pi\nu\tau} \delta(\vec{x} - \vec{x}^{\ \prime}) \quad ,
\label{eqb1}
\end{equation}
where $\nu$ is the local density of states and $\tau$ the elastic mean free
time (we put $\hbar = m =1$).

With $\tau=l_{el}/k_F$ and $N=(k_FL/\pi)^d$ in $d$ dimensions we get for the
total density of states
\begin{equation}
\rho = {dN\over dE} = {d\over 2} k_f^{d-2} 
\left( {L\over \pi} \right)^d
\label{eqb2}
\end{equation}
and hence
\begin{equation}
\nu = {\rho\over L^d} = {d\over 2} k_F^{d-2} \pi^{-d}
\label{eqb3}
\end{equation}
so that
\begin{equation}
c_d = {1\over 2\pi\nu\tau} = {\pi^{d-1}\over d} {k_f^{4-d}\over k_Fl_{el}} 
\quad ,
\label{eqb4}
\end{equation}
which is exactly (\ref{eq29}).

\section{Final Evaluation of $1/\xi$}

\label{app:axel_c}

We have to estimate the r.h.s. of (\ref{eq35}). Using (\ref{eq8}) and
(\ref{eq34}) an elementary calculation shows that
\begin{equation}
B_{nm} = 
{1\over 2}
\sum_{s_1,s_2 = \pm 1}
{ \sin \left[ (\kappa_n+\kappa_m)L/2 + s_1\pi(n+1/2) + s_2\pi(m+1/2) \right]
  \over
              (\kappa_n+\kappa_m)L/2 + s_1\pi(n+1/2) + s_2\pi(m+1/2) }
\quad .
\label{eqc1}
\end{equation}
With the abbreviation $f(x)=\sin(x)/x$ (\ref{eq35}) can therefore be rewritten
as
\begin{eqnarray}
{1\over \xi} &=& 
{1\over l_{el}} \sum_{s_1\ldots s_4} \sum_{nm} {1\over \eta_n\eta_m}
f\left( \pi\left[ \eta_n+\eta_m + s_1(n+1/2) + s_2(m+1/2) \right] \right)
\nonumber\\
& &\phantom{
{1\over l_{el}} \sum_{s_1\ldots s_4} \sum_{nm} {1\over \eta_n\eta_m}
}
f\left( \pi\left[ \eta_n+\eta_m + s_3(n+1/2) + s_4(m+1/2) \right] \right)
\quad .
\label{eqc2}
\end{eqnarray}
With the help of the approximation scheme
\begin{eqnarray}
\sum_{n=0}^{N-1} {1\over \eta_n} h(n+1/2) &\approx&
\int_0^N {dx\over\sqrt{N^2-x^2}} h(x)   \nonumber\\
&=& \int_0^1 {ds\over\sqrt{1-s^2}} h(Ns) \nonumber\\
&=& \int_0^{\pi/2} d\varphi \ h(N\sin\varphi)
\label{eqc3}
\end{eqnarray}
for an arbitrary function $h(n+1/2)$ we find
\begin{eqnarray}
{1\over\xi} &=&
{1\over l_{el}} \sum_{s_1\ldots s_4} \int_0^{\pi/2} d\varphi_1 d\varphi_2 \
f\left(\pi N\left[ \cos\varphi_1 + \cos\varphi_2 + s_1 \sin\varphi_1
                  + s_2\sin\varphi_2 \right] \right)    \nonumber\\
& &\phantom{
{1\over l_{el}} \sum_{s_1\ldots s_4} \int_0^{\pi/2} d\varphi_1 d\varphi_2
}
f\left(\pi N\left[ \cos\varphi_1 + \cos\varphi_2 + s_3 \sin\varphi_1
                  + s_4\sin\varphi_2 \right] \right)  
\quad .
\label{eqc4}
\end{eqnarray}
The main contributions to the integral arise from the regions where
$[\ldots ] \approx 1/N$. Also, the two factors in the integrand should 
be in phase,
i.e. $s_1=s_3$ and $s_2=s_4$ (diagonal approximation). Therefore we are 
left with
the expression (using $\sin\varphi+\cos\varphi = \sin(\varphi+\pi/4)/\sqrt{2}$)
\begin{equation}
{1\over\xi} = {1\over l_{el}} \int_{-\pi/2}^{\pi/2}
d\varphi_1 d\varphi_2 \
f^2 \left( {\pi N\over \sqrt{2}} \left[ 
   \sin(\varphi_1+\pi/4) + \sin(\varphi_2+\pi/4) \right] \right) 
\quad ,
\label{eqc5}
\end{equation}
for which the integration range can be restricted to a narrow strip around the
line $\varphi_1 = -\pi/2-\varphi_2$. A first order Taylor expansion in the
deviation $\Delta\varphi$ from this line yields
\begin{equation}
{1\over\xi}
\approx
{1\over l_{el}}
\int_{-\pi/2}^0 d\varphi \int_{-infty}^\infty d(\Delta\varphi) \
f^2\left( {\pi N\over\sqrt{2}} \cos(\varphi+\pi/4) \Delta\varphi\right) \quad .
\label{eqc6}
\end{equation}
With the help of the integral
\begin{equation}
\int_{-\infty}^\infty dx\ {\sin^2(ax) \over (ax)^2} = {\pi\over |a|}
\label{eqc7}
\end{equation}
we get
\begin {eqnarray}
{1\over \xi} &=& {\pi\over l_{el}} \int_{-\pi/2}^0 d\varphi \
{\sqrt{2}\over \pi N} {1\over |\cos(\varphi+\pi/4)|} \nonumber\\
&=&
{\sqrt{2}\over Nl_{el}} \int_{-\pi/4}^{\pi/4} {d\varphi\over\cos\varphi}
\nonumber\\
&=&
{\sqrt{2}\over Nl_{el}} \ln{\tan(3\pi/8)\over\tan(\pi/8)}
= 
{2\sqrt{2}\over Nl_{el}} \ln(1+\sqrt{2}) \quad ,
\label{eqc8}
\end{eqnarray}
which is exactly (\ref{eq36}).

\eject



\noindent


\begin{thebibliography}{99}

\bibitem{thouless} D. C. Thouless, Phys.\ Rev.\ Lett.\ {\bf 39}, 1167 (1977).

\bibitem{scaling1} E. Abrahams, P. W. Anderson, D. C. Licciardello, and 
  T. V. Ramakrishnan, Phys. Rev. Lett. {\bf 42}, 673 (1979). 

\bibitem{scaling2} P. W. Anderson, D. J. Thouless, E. Abrahams, and 
  D. S. Fisher, Phys. Rev. B {\bf 15}, 3519 (1980).

\bibitem{wegner1} F. J. Wegner, Z. Physik B {\bf 35}, 207 (1979).

\bibitem{efetov} K.B. Efetov, Adv. in Phys. {\bf 32}, 53 (1983).

\bibitem{wegner2} F. J. Wegner, Nucl. Phys. B {\bf 316}, 663 (1989). 

\bibitem{hikami} S. Hikami, Prog. Theor. Phys. {\bf 107}, 213 (1992). 

\bibitem{larkin} K. B. Efetov and A. I. Larkin, Zh. Eksp. Theor. Fiz. 
        {\bf 85}, 764 (1983); Sov. Phys. JETP {\bf 58}, 444 (1983).

\bibitem{zirn_cond} M. R. Zirnbauer, Phys. Rev. Lett. {\bf 69}, 
  1584 (1992). 

\bibitem{mmz} \label{mmz} A. D. Mirlin, A. M\"uller-Groeling, and 
        M. R. Zirnbauer, Ann. Phys. (N.Y.) {\bf 236}, 325 (1994).


\bibitem{fokpla} O. N. Dorokhov, Pis'ma Zh. Eksp. Teor. Fiz.
        {\bf 36}, 259 (1982) [JETP Lett. {\bf 36}, 318 (1982)]; 
        P. A. Mello, P. Pereyra, and. N. Kumar, Ann.
        Phys. (N. Y.) {\bf 181}, 290 (1988).

\bibitem{GUEcase} C. W. J. Beenakker and B. Rejaei, Phys. Rev. Lett. 
        {\bf 71}, 3689 (1993); C. W. J. Beenakker and B. Rejaei, 
        Phys. Rev. B {\bf 49}, 7499 (1994). 

\bibitem{frahm_let} K. Frahm, Phys. Rev. Lett. {\bf 74}, 4706 (1995).

\bibitem{eq_rejaei} B. Rejaei, Phys. Rev. B {\bf 53}, R13235 (1996).

\bibitem{equiv} P. W. Brouwer and K. Frahm, Phys. Rev. B
  {\bf 53}, 1490 (1996).

\bibitem{shep1} D. L. Shepelyansky, Phys. Rev. Lett. {\bf 73}, 2607 (1994).

\bibitem{imry} Y. Imry, Europhys. Lett. {\bf 30}, 405 (1995).

\bibitem{fmgpw} K. Frahm, A. M\"uller--Groeling, J.- L. Pichard, and
  D. Weinmann, Europhys. Lett. {\bf 31}, 169 (1995).

\bibitem{wmgpf} D. Weinmann, A. M\"uller--Groeling, J.--L. Pichard, and 
        K. Frahm, Phys. Rev. Lett. {\bf 75}, 1598 (1995).

\bibitem{oppen1} F. von Oppen, T. Wettig, and J. M\"uller, Phys. Rev. Lett. 
  {\bf 76}, 491 (1996).


\bibitem{shep3} F. Borgonovi and D. L. Shepelyansky, 
        Nonlinearity {\bf 8}, 877 (1995). 

\bibitem{shep4} F. Borgonovi and D. L. Shepelyansky, 
        J. Phys. I France {\bf 6}, 287 (1996).

\bibitem{fmgp} K. Frahm, A. M\"uller--Groeling, and J.- L. Pichard, Phys. Rev. 
Lett. {\bf 76}, 1509 (1996).

\bibitem{dorokhov} O. N. Dorokhov, Zh. Eksp. Teor. Fiz. {\bf 98}, 646
  (1990) [Sov. Phys. JETP {\bf 71}, 360 (1990)].

\bibitem{ip} Y. Imry and J.- L. Pichard, Saclay preprint.

\bibitem{shep2} P. Jacquod and D. L. Shepelyansky, Phys. Rev. Lett. 
        {\bf 75}, 3501 (1995). 

 \bibitem{fyodorov2} Y. V. Fyodorov and A. D. Mirlin, Phys. Rev. B {\bf
    52}, R11580 (1995).

\bibitem{framg1} K. Frahm and A. M\"uller--Groeling, Europhys. Lett. 
        {\bf 32}, 385 (1995). 



\bibitem{vwz} \label{vwz} J. J. M. Verbaarschot, H. A. Weidenm\"uller, and M.
R. Zirnbauer, Phys. Rep. {\bf 129}, 367 (1985).

\bibitem{iwz} \label{iwz} S. Iida, H.A. Weidenm\"uller, and J. A. Zuk, Ann.
Phys. (N.Y.) {\bf 200}, 219 (1990).
 

\bibitem{fyodorov} Y. V. Fyodorov and A. D. Mirlin, Phys. Rev. Lett. 
        {\bf 67}, 2405 (1991); ibid. {\bf 69}, 1093 (1992);
        ibid. {\bf 71}, 412 (1993). 

\bibitem{mehta} M. L. Mehta, {\em Random Matrices\/}, 2nd ed. 
        (Academic, New York, 1991).

\bibitem{alt1} B. L. Al'tshuler and B. I. Shklovskii, Zh. Eksp. Teor. 
        Fiz. {\bf 91}, 220 (1986) [Sov. Phys. -- JETP {\bf 64}, 127 (1986)].




\bibitem{oppen2} F. von Oppen and T. Wettig, Europhys. Lett. {\bf 32}, 
  741 (1995). 


\bibitem{akpi} E. Akkermans and J.--L. Pichard, preprint (1996).


\bibitem{pc}
A. M\"uller--Groeling, H.A. Weidenm\"uller and C.H. Le\-wen\-kopf, Europhys. Lett.
{\bf 22}, 193 (1993);
A. M\"uller--Groeling and H.A. Weidenm\"uller, Phys. Rev. {\bf B 49}, 4752 (1994).




















                    

 


 

 
 


\end{thebibliography}
\end{document}